\documentclass[aps,prb,twocolumn]{revtex4-1}

\usepackage{amsmath} 
\usepackage{amsthm} 
\usepackage{graphicx} 
\usepackage{float}
\usepackage{amssymb}
\usepackage{enumitem}

\begin{document}

\title{Tuning current flow in superconducting thin film strips by control wires. Applications to single photon detectors and diodes. }

\author{Alex Gurevich}
\email[]{gurevich@odu.edu}
\affiliation{Department of Physics, Old Dominion University, Norfolk, Virginia 23529, USA.}

\date{\today}

\begin{abstract}
It is shown that integration of a thin film superconducting strip with current-carrying control wires enables one to engineer a profile of supercurrent density $J(x)$ with no current crowding at the edges of a strip wider than the magnetic Pearl length $\Lambda$. Moreover, $J(x)$ in a strip can be tuned by control wires to produce an inverted $J(x)$ profile with  dips at the edges to mitigate current crowding at lithographic defects and block premature penetration of vortices. These conclusions are corroborated by calculations of $J(x)$ in a thin strip coupled inductively with side control wires or in bilayer strip structures by solving the London and Ginzburg Landau equations in the thin film Pearl limit. Thermally-activated penetration of vortices from the edges and unbinding of vortex-antivortex pairs in inverted $J(x)$ profiles are evaluated. It is shown that these  structures can be used to develop single-photon strip detectors much wider than $\Lambda$. Such detectors can be tuned {\it in situ} by varying current in control wires to reach the ultimate photon sensitivity limited by unbinding of vortex-antivortex pairs. The structures considered here exhibit a non-reciprocal current response and behave as superconducting diodes. They can also be used to study the physics of vortex matter in thin films not masked by penetration of vortices from the edges. 

\end{abstract}

\maketitle

\section{Introduction}\label{section_introduction}
Thin superconducting films and micro resonators are instrumental in superconducting electronics ~\cite{res,qq}. Particularly, superconducting nanowire single-photon detectors (SNSPD)~ \cite{spd1,spd2,spd3} with up to ps time resolution have proliferated in quantum information technology ~\cite{qi1,qi2,qi3,qi4,qi5,qi6,qi7,qi8}, astronomy ~\cite{astron}, high energy particle detectors and  search for dark matter ~\cite{hep1,hep2,hep3}, or deep brain imaging ~\cite{brain1,brain2,brain3}. One of the goals of the SNSPD technology is the development of large-area detector arrays with a nearly 100 $\%$  photon sensitivity into infrared light ~\cite{spd2,spd3,brain3}. Such detectors are photon-sensitive at currents $I_i<I_s<I_b$, where the onset current $I_i$ and the switching current $I_b$ are fractions of the depairing current $I_d$. 

\begin{figure}[h]
   	\includegraphics[scale=0.45,trim={0mm 0mm 0mm 0mm},clip]{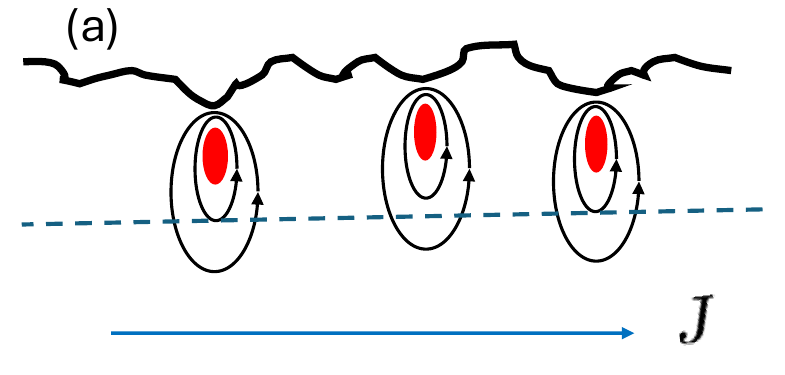}
	\includegraphics[scale=0.42,trim={0mm 0mm 0mm 0mm},clip]{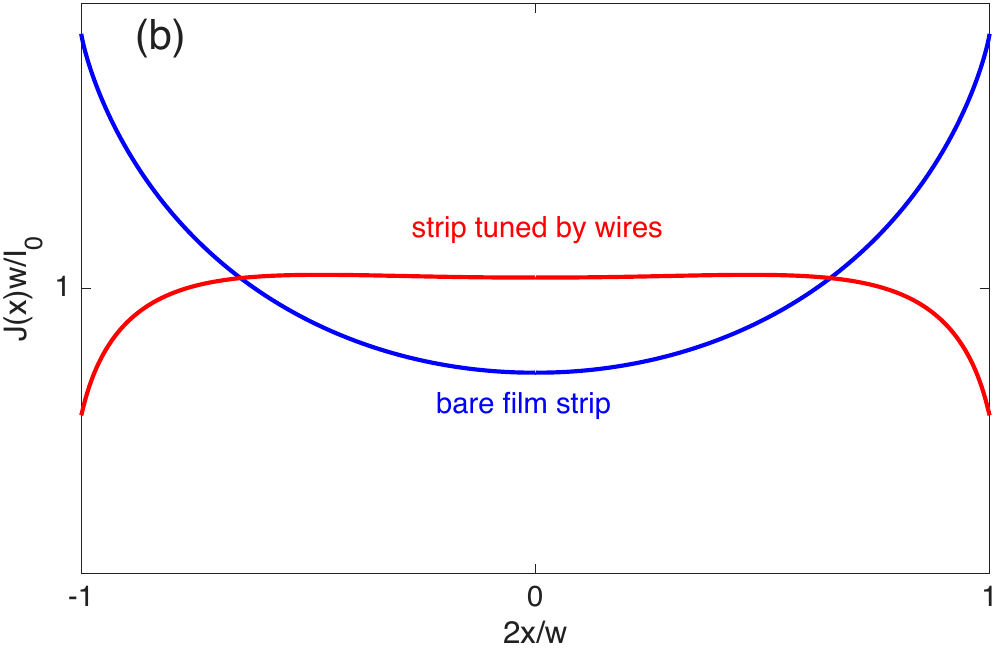}
   	\caption{(a) Edge roughness causing local current crowding and premature penetration of vortices. The dashed line depicts a boundary of a layer in which variations of ${\bf J}({\bf r})$ occur on the scale of the edge roughness assumed much smaller than the film width and $\Lambda$. (b) The Pearl current crowding at the edges of a bare strip (blue). An inverted sheet current density profile $J(x)$ in a strip tuned by control wires (red) deactivating both the edge defects and the Pearl current crowding.  }
   	\label{F1}
   	\end{figure}

Boosting the sensitivity of SNSPD requires increasing both the operational current range $I_b-I_i$ and the detector area. As the bias current $I$ approaches $I_d$, the photon energy required to form a critical hotspot expanding across the strip decreases and the minimum detectable photon energy becomes weakly dependent on the strip width ~\cite{vodolaz} so it would be beneficial to push $I_b$ toward $I_d$ as much as possible. However, there are limitations of how close $I_b$ can approach $I_d$ and how much the width of a straight thin film strip be increased. First, thin strips inevitably have lithographic defects causing localized current crowding along the edges~ \cite{def1,def2}. For instance, a semi-circular edge indentation of radius larger than the coherence length $\xi$ causes a two-fold local increase in the sheet current density $J(x)$, while a sharp indentation or microcrack causes even greater current crowding which can initiate vortex jets ~\cite{jet1,jet2,jet3,jet4} or dendritic thermo-magnetic vortex avalanches ~\cite{dendr1,dendr2,dendr3} destroying the non-dissipative current-carrying state.  Second, the Meissner effect causes current crowding at the edges  which becomes detrimental for the SNSPD performance as the width of the strip exceeds the magnetic Pearl length $\Lambda$~ \cite{pearl,dg}. Interplay of these effects illustrated in Fig. \ref{F1} decreases the entry energy barrier of vortices, greatly increasing the dark count rate due to thermally-activated penetration of vortices from the edges at currents well below $I_d$.  To increase the photon-sensitive area, meandering SNSPDs structures have been developed  ~\cite{spd1,spd2,spd3}. To mitigate the current crowding by the edge roughness, the active SNSPD areas have been irradiated with heavy ions leaving intact the long edges along  the strip ~\cite{irad1} and the bend regions ~ \cite{irad2}.   

This work explores boosting the performance of SNSPD and other thin film nanostructures by forming an inverted profile of $J(x)$ in a superconducting strip shown in Fig. \ref{F1}. Here a reduced current density at the edges $J_e$ mitigates the effect of lithographic defects and eliminates current crowding caused by the Pearl screening. It is shown that such inverted profiles of $J(x)$ can indeed be produced and {\it tuned in situ} in a superconducting strip coupled inductively with adjacent current-carrying wires or other thin film structures. These control wires produce the magnetic field that counters the perpendicular component of the self field of SNSPD and redistribute $J(x)$ in the strip as shown in Fig. \ref{F1}. Such structures  can provide a flat portion of $J(x)$ in the photon-sensitive area of the strip and controllable dips at the edges which block premature penetration of vortices and give rise to a non-reciprocal current-voltage characteristic of a superconducting diode. Furthermore, control currents can be tuned in such a way that the SNSPD reaches the ultimate performance limit determined by the unbinding of vortex-antivortex pairs in the strip ~\cite{bkt1,bkt2}. Such device architecture lifts the SNSPD size limitations and opens  opportunities for the development of straight strip detectors much wider than the Pearl length. This approach has been recently implemented in Ref. \onlinecite{natp} which reported the development of 3 nm thick and up to 0.1 mm wide WSi strip detectors integrated with Nb side wires ("rails"), reaching 100$\%$ sensitivity into infrared, up to 20 $\%$ larger switching currents and the dark count rate decreased by up to 9 orders of magnitude.  Manipulation of resistive switching in superconducting films by current-carrying control wires has a long history which goes back to the development of cryotrons ~\cite{cr0,cr1,cr2,cr3}. 
 
The paper is organized as follows.  In Sec. II $J(x)$ in a thin film strip coupled with control wires is calculated by solving the London equations. It is shown that the inverted profile of $J(x)$ can be produced in strips of any width including strips much wider than the Pearl length. In Sec. III bilayers SNSPD structures are considered in which one of the layers is used to tune $J(x)$ in the prime strip. In Sec. IV the nonlinear pairbreaking effects in a bare strip and a strip with control wires are considered by solving the Ginzburg-Landau (GL) equations. It is shown that the strip inductively coupled with control wires has a non-reciprocal current response and can behave as a tunable superconducting diode. In Sec. V the dynamics of vortices driven by the inverted profile of $J(x)$ is considered. The transition from the dark counts rate caused by penetration of vortices through the edges to unbinding of vortex-antivortex pairs in the bulk upon increasing the control current is calculated. Sec. VI concludes with discussions of the results.      

\section{A strip between control wires} \label{S1}

\begin{figure}[h]
   	\includegraphics[scale=0.6,trim={0mm 0mm 0mm 0mm},clip]{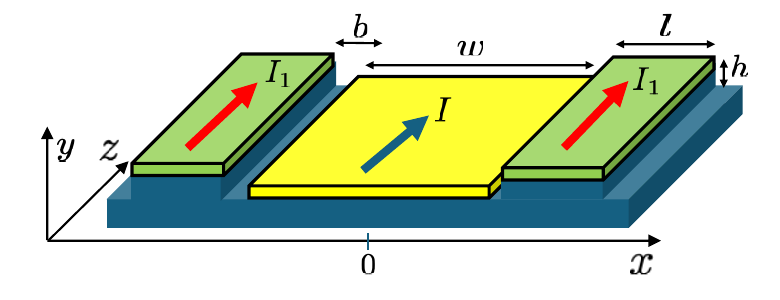}
   	\caption{A strip of width $w$ between two control strip wires of width $l$. The coordinate axes are shifted downward for clarity.}
   	\label{F2}
   	\end{figure}

Consider a long thin strip of width $w$ in the plane $y=0$ between two control strip wires of width $l$ at $w/2+b<x<w/2+b+l$ and $-w/2-b-l<x<-w/2-b$, as shown in Fig. \ref{F2}. Each control wire is raised by $h$ and spaced by $b$ from the edge of the central strip. Here the central strip carries a dc current $I$ and each control wire carries a dc current $I_1$, all strips are infinite along $z$ and no external magnetic field is applied. 

The magnetic field $B_z$ along the infinite strip vanishes and $B_x=\partial A/\partial y$ and $B_y=-\partial A/\partial x$ are expressed in terms of the vector potential ${\bf A}(x,y)=[0,0,A(x,y)]$ caused by currents flowing along $z$. In this paper thin films of thickness $d$ much smaller than the London penetration depth $\lambda$ are considered for which $A(x,y)$ satisfies the London-Maxwell equation~ \cite{pearl,dg,ehb}:
\begin{equation}
\lambda^2\nabla^{2}A-dQ(x)\delta(y)=0,
\label{lond}
\end{equation}
where the gauge-invariant ${\bf Q}={\bf A}+(\phi_{0}/2\pi)\nabla\theta$ is proportional to the current density ${\bf j}=-{\bf Q}/\mu_0\lambda^2$ in the film,  $\theta$ is the phase of the superconducting order parameter and $\phi_0$ is the magnetic flux quantum. The boundary condition $B_x(x,+0)-B_x(x,-0)=\mu_0J(x)$, where $J(x)=\int j(x,y)dy$ is the sheet current density readily follows from Eq. (\ref{lond}). Furthermore,  $j_x=j_y=0$ and $j_z(x)$ flowing along the strip can depend only on $x$ so the phase gradient $\nabla\theta=(0,0,\theta')$  has only a constant z-component $\theta'$ independent of $x$. The solution of Eq. (\ref{lond}) for $A(x,y)$ around the strip is given by: 
\begin{gather} 
A(x,y)=\frac{dw}{8\pi\lambda^{2}}\int_{-1}^{1}\ln[(x-u)^{2}+y^{2}]Q(u)du+
\label{A1} \\ 
\frac{d_{1}w}{8\pi\lambda_{1}^{2}}\biggl[\int_{1+b}^{1+b+l}\!\!\!\ln[(x-u)^{2}+(y-h)^{2}]Q_{1}(u)du+
\\ \nonumber
\int_{-1-b-l}^{-1-b}\!\!\!\ln[(x-u)^{2}+(y-h)^{2}]Q_{2}(u)du\biggr].
\nonumber
\end{gather}
Here $Q(x)$, $Q_1(x)$ and $Q_{2}(x)$ describe superflow in the strip in the absence of vortices, the right and the left control wires, respectively, $\lambda$ and $\lambda_1$ are the London penetration depths in the strip and the control wire, $d$ and $d_1$ are their thicknesses, and all lengths are in units of $w/2$. If both control wires carry equal parallel currents, $Q(u)=Q(-u)$, $Q_{1}(u)=Q_{2}(-u)$ and Eq. (\ref{A1}) becomes:
\begin{gather} 
A(x,y)=k\int_{0}^{1}\ln[(u^2-x^2+y^2)^2+4x^2y^2]Q(u)du+
\nonumber \\
k_1\int_{1+b}^{1+b+l}\ln[(x^2-u^2)^2+2(y-h)^2(x^2+u^2)+
\nonumber \\
(y-h)^4]Q_{1}(u)du,
\label{A2} 
\end{gather}
where $k=w/4\pi\Lambda$, $k_1=w/4\pi\Lambda_1$, and $\Lambda=2\lambda^2/d$ and $\Lambda_1=2\lambda_1^2/d_1$ are the Pearl screening lengths ~\cite{pearl} in the strip and control wires, respectively. 

\begin{figure}[h]
   	\includegraphics[scale=0.48,trim={5mm 0mm 0mm 0mm},clip]{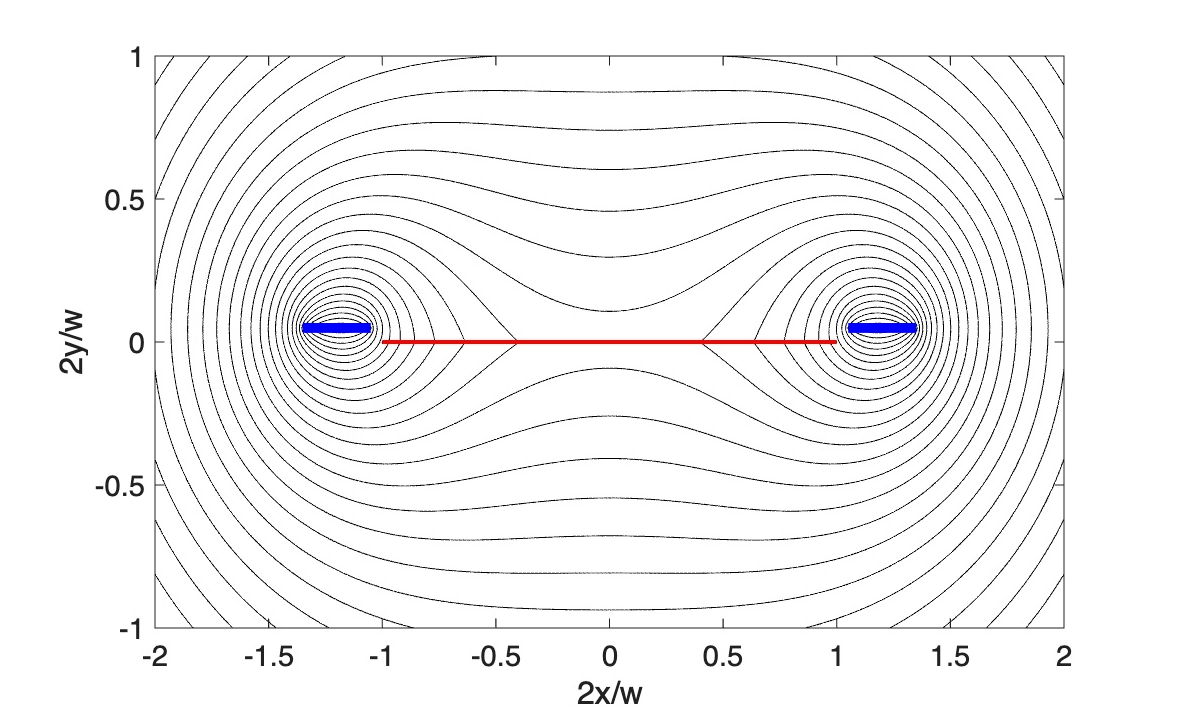}
   	\caption{Magnetic field lines calculated from Eqs. (\ref{A2}) - (\ref{Q1}) for $l=0.15w$, $k=0.03$, $k_1=50k$, $b=h=0.05w$.}
   	\label{F3}
   	\end{figure}

Setting $y=0$ in Eq. (\ref{A2}) and using $A(x,0)=Q(x,0)-\phi_0\theta'/2\pi$ gives two equations for $Q(x)$ and $Q_1(x)$ ~\cite{natp}:
\begin{gather}
Q(x)-2k\int_0^1\!\ln|x^{2}-u^{2}|Q(u)du-
\label{Q0} \\
k_{1}\int_{1+b}^{1+b+l}\!\!\ln[(u^{2}+h^{2}-x^{2})^{2}+4h^{2}x^{2}]Q_{1}(u)du=\alpha,
 \nonumber \\
Q_1(x)-2k_{1}\int_{1+b}^{1+b+l}\!\!\ln|x^{2}-u^{2}|Q_{1}(u)du-
\label{Q1} \\
k\int_0^1\!\ln[(u^{2}+h^{2}-x^{2})^{2}+4h^{2}x^{2}]Q(u)du=\beta,
\nonumber
\end{gather}
where $\alpha=\phi_{0}\theta_{0}'/2\pi$ and $\beta=\phi_0\theta_1'/2\pi$ with constant phase gradients $\theta'_0$ and $\theta_1'$.
Equations (\ref{Q0}) and (\ref{Q1}) were solved as described in Appendix A, where $\alpha$ and $\beta$ are expressed in terms of currents $I$ and $I_1$~\cite{natp}. Tuning of $J(x)$ is most effective if the control wires carry much higher current densities than the central strip, which requires $k_1\gg k$. This case is illustrated here for a 4 nm thick W$_{0.8}$Si$_{0.2}$ amorphous film with $T_c=4.1$ K, $\xi=7$nm,  $\lambda=700$ nm and $\Lambda=245\,\mu$m ~\cite{wsi} at 1K in the strip, and a moderately clean Nb with $\lambda=50$ nm, $\xi=32$ nm in the control wires.  For these parameters, the ratio $k_1/k$ varies from 200 to 1000 as $d_1$ is increased from 4 nm to 20 nm but $k_1/k$ can be lower for higher concentration of nonmagnetic impurities in Nb. After $Q(x)$ and $Q _1(x)$ are obtained, ${\bf H}(x,y)$ around the strip is calculated using Eq. (\ref{A1}). As shown in Fig. \ref{F3}, the magnetic field of control wires counters the self field of the strip at the edges, mitigating the Pearl current crowding.

\begin{figure}[h]
   	\includegraphics[scale=0.4,trim={0mm 0mm 0mm 0mm},clip]{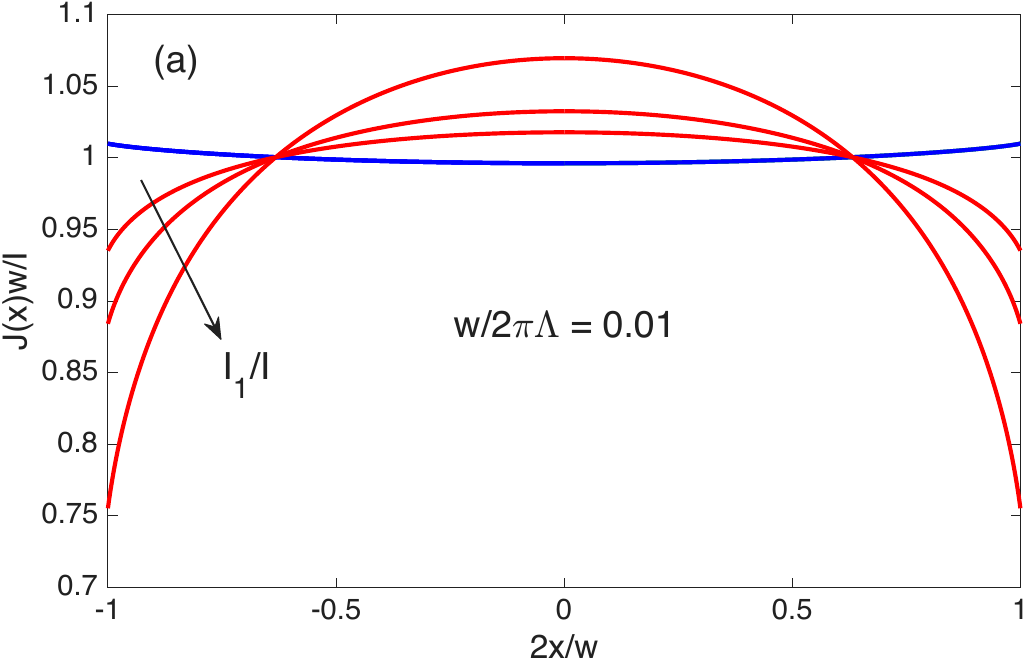}
	\includegraphics[scale=0.4,trim={0mm 0mm 0mm 0mm},clip]{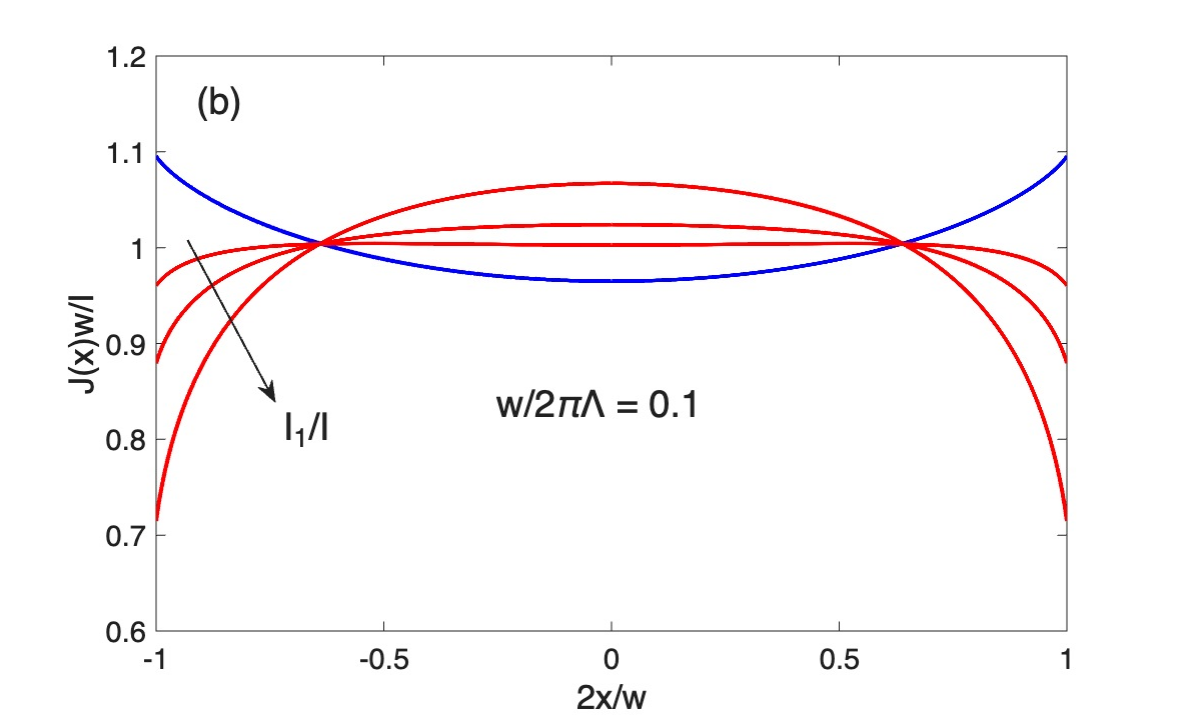}
	\includegraphics[scale=0.4,trim={5mm 80mm 0mm 75mm},clip]{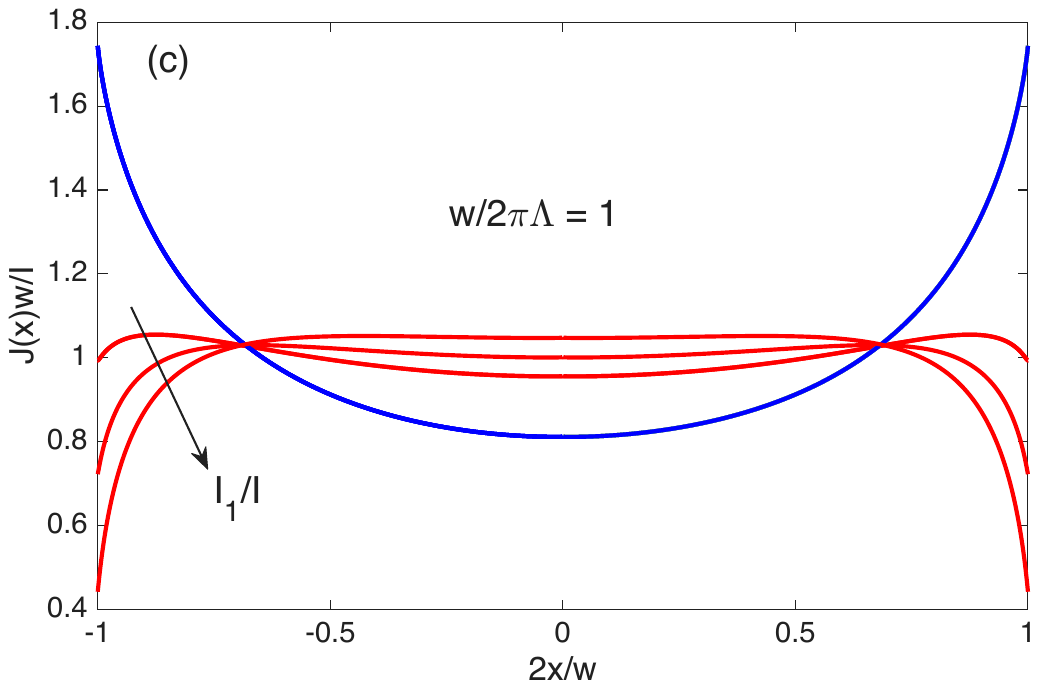}
   	\caption{Sheet current density calculated for different strip widths and control currents, $k_1=500k$, $l=0.15w$, $b=0.005w$ and $h=0$:  (a) $w/2\pi\Lambda=0.01$, $I_1/I=3.05, 5.11, 10.3$, (b) $w/2\pi\Lambda=0.1$, $I_1/I=0.61, 0.94, 1.59$, (c) $w/2\pi\Lambda=1$, $I_1/I=0.45, 0.58, 0.71$. For W$_{0.8}$Si$_{0.2}$ films with $d=4$ nm and  $\Lambda = 245,\mu$m ~\cite{wsi}, Figs. 4a-c correspond to $w\approx 15.4\, \mu$m, $154\, \mu$m and $1.54$ mm, respectively.  The blue lines show $J(x)w/I$ for a single strip.}
   	\label{F4}
   	\end{figure}

Examples of $J(x)$ in strips of different widths calculated from Eqs. (\ref{Q0}) and (\ref{Q1}) are shown in Fig. \ref{F4}. Here the blue lines show the results for single strips in which $J(x)$ peaks at the edges due to the Pearl screening, the peaks increasing as the strip gets wider. Raising the control current $I_1$ eliminates the Pearl peaks in $J(x)$ and produces inverted $J(x)$ profiles with dips at the edges which deepen as $I_1$ is increased. For the above parameters, the dips in $J(x)$ are produced by control currents well below the depairing current, while the local $J(x)$ at the edge or the center of the strip can approach $J_d$, as will be shown below. For a narrow strip with $w< 0.1\Lambda$ represented by Fig. \ref{F4}a, the self-field effects are weak, so the control current $I_1$ is to be large enough $\simeq (2-10)I$ to push the current density in the strip away from the edges.  As the width increases, the dips in $J(x)$ at the edges can be produced by smaller control currents relative to $I$. Shown in Fig. \ref{F4}c is $J(x)$ in a strip more than 6 times wider than $\Lambda$. The evolution of $J(x)$ with $w$ shown in Fig. \ref{F4} is controlled by the ratios of $w/\Lambda$, $l/w$ and $\Lambda_1/\Lambda$, irrespective of specific materials parameters. 

The London theory not only captures the essential physics of tuning $J(x)$ by control wires but also adequately describes $J(x)$ in SNSPD operating at large currents $I\sim I_d$. This is addressed in Sec. IV, where it is shown that $J(x)$ obtained from the London theory is nearly identical to $J(x)$ obtained from the GL theory taking into account a weak dependence of $\Lambda$ on $J$ at $I\leq 0.8I_d$. Another issue is that the control wires are to remain in the superconducting states as $I_1$ is increased. A qualitative estimate showing that it is indeed the case can be obtained using $I= Qw/2\mu_0\Lambda$ and $I_1= Q_1l/2\mu_0\Lambda_1$, where $Q$ and $Q_1$ are mean values of $Q(x)$ and $Q_1(x)$. Hence, $Q_1/Q= (\Lambda_1w/\Lambda l)(I_1/I)=I_1/75I$ for $\Lambda/\Lambda_1=500$ and $l=0.15w$ used to produce the results shown in Fig. \ref{F4}.  Even if $Q$ in the strip is close to the GL pairbreaking limit $Q_c=\phi_0/2\pi\sqrt{3}\xi$, the control phase gradient $Q_1$ remains below $Q_{c1}=\phi_0/2\pi\sqrt{3}\xi_1$. For instance, taking $Q=\eta Q_c$ with $\eta=0.8$ gives $Q_1/Q_{c1}\simeq \eta I_1\xi_1/75\xi I\simeq 0.05I_1/I$ at $\xi=7$ nm and $\xi_1=32$ nm. Thus, $Q_1<Q_{c1}$ for all cases  shown in Fig. \ref{F4}, particularly in wider strips for which $I_1\simeq I$.  Calculation of $Q_1(x)$ taking into account the Pearl current crowding in the control wire and the GL current pairbreaking in the central strip presented in Sec. IV is consistent with the above conclusion. 

Tuning $J(x)$ in the strip by control current eliminates the Pearl current crowding in a strip of any width while producing controllable dips in $J(x)$ at the edges. The latter deactivate the lithographic defects depicted in Fig. \ref{F1}a by reducing $J(\pm w/2)$ and mitigating the current crowding around edge defects.  The extent of current crowding by edge defects is sample-dependent and is a-priori unknown. Yet  tuning $J(x)$ in-situ can reduce $J(\pm w/2)$ to any desirable level at which the resistance of a strip is no longer controlled by penetration of vortices from the edges but is determined by unbinding of vortex-antivortex pairs in the strip, as shown below.

\section{Bilayers} \label{S2}
This chapter addresses tuning $J(x)$ in bilayers in which two inductively-coupled strips carry antiparallel currents. Such bilayers can be used for the development of ultra-wide SNSPD with no Pearl current crowding and as building blocks in SNSPD arrays in  multi-pixel single photon cameras.  Two cases are considered: 1. The top and the bottom films in the bilayer are identical and electrically connected at one end. 2. The top and the bottom films made of different superconductors are disconnected and carry different currents.  

\subsection{Connected bilayers}
\begin{figure}[h!]
   	\includegraphics[scale=0.38,trim={0mm 10mm 0mm 10mm},clip]{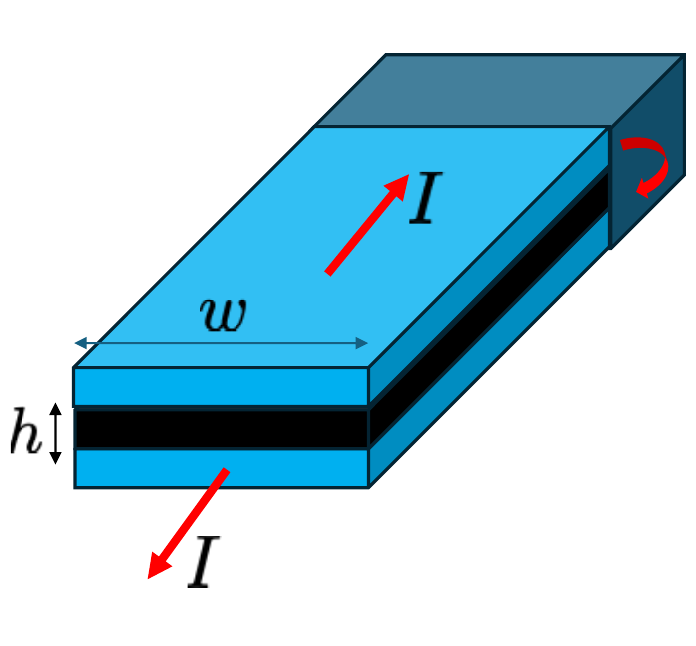}
   	\caption{Two long thin film strips electrically connected at one end and carrying equal antiparallel currents $I$. A dielectric interlayer of thickness $h$ is shown in black.}
   	\label{F5}
   	\end{figure}

Consider a bilayer with two stacked identical films separated by a dielectric layer of thickness  $h$ and electrically connected at one end, as shown in Fig. \ref{F5}. The films carry antiparallel currents $I$ injected into one of the films.  The dielectric layer is assumed to be thick enough to suppress Josephson coupling causing phase textures ~ \cite{gv}.

\begin{figure}[h!]
   	\includegraphics[scale=0.42,trim={0mm 0mm 0mm 0mm},clip]{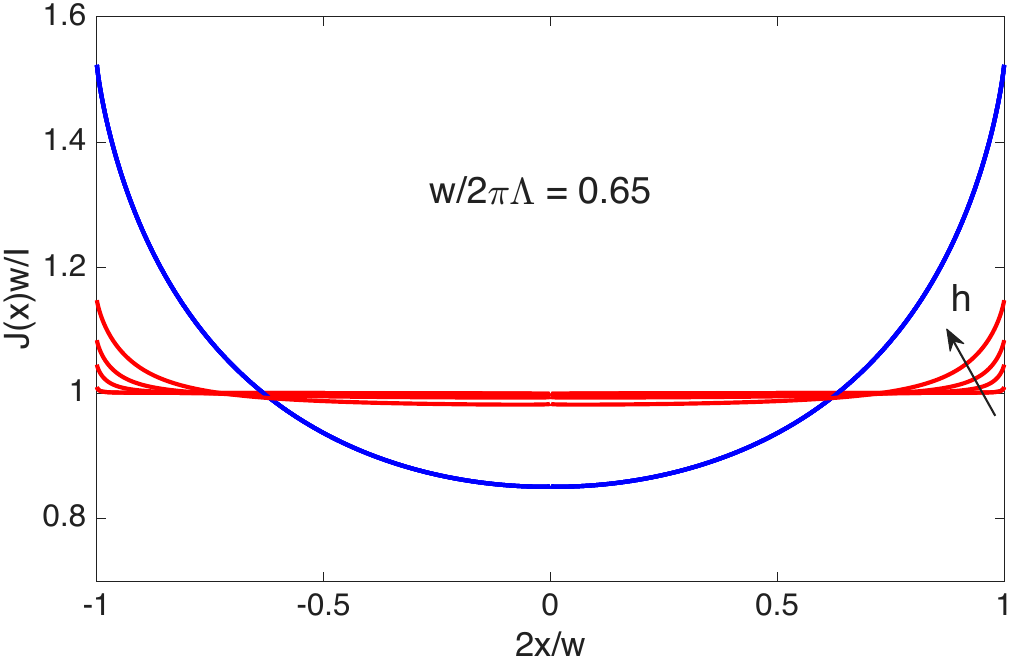}
   	\caption{$J(x)$ in a bifilar bilayer calculated at $w/2\pi\Lambda = 0.65$  ($w=1$ mm for W$_{0.8}$Si$_{0.2}$) and interlayer spacings: $h/w=0.005, 0.025, 0.05, 0.1$. For these $h$, the normalized current density at the edge $J_ew/I$ reaches $1.0095, 1.045, 1.084, 1.15$, respectively. The blue line shows $J(x)$ in a single strip.  }
   	\label{F6}
   	\end{figure}

In this geometry $Q(x)$ in the top film and $Q_1(x)$ in the bottom films are related by $Q(x)=-Q_1(x)$, so the solution of Eq. (\ref{lond}) takes the form:
\begin{eqnarray}
A(x,y)=\frac{dw}{8\pi\lambda^{2}}\int_{-1}^{1}\ln[(x-u)^{2}+y^{2}]Q(u)du-
\label{Ast} \\
\frac{dw}{8\pi\lambda^{2}}\int_{-1}^{1}\ln[(x-u)^{2}+(y+h)^{2}]Q(u)du
\nonumber
\end{eqnarray}
Setting here $A(x,0)=Q(x)-\phi_0\theta_0'/2\pi$ at $y=0$ yields the self-consistency equation for $Q(x)$:
\begin{equation}
\!\!Q(x)-k\!\int_{0}^{1}\!\ln\!\left[\frac{(x^{2}-u^{2})^{2}}{(x^{2}+h^{2}-u^{2})^{2}+4h^{2}u^{2}}\right]\!Q(u)du=\alpha,
\label{Q2}
\end{equation}
where $k=w/4\pi\Lambda$ and $\alpha=\phi_0\theta_0'/2\pi$. The 
sheet current densities $J=2Q(x)/\mu_0\Lambda$ calculated for different layer spacings $h$ are shown in Fig. \ref{F6}.
Here $J(x)$ is nearly flat because two strips with equal antiparallel currents produce no magnetic field far away from the bilayer.  
Small peaks in $J(x)$ at the edges result from the field leakage from the interlayer space, the peaks decrease as $h$ is decreased.  The Pearl current crowding practically disappears in a bifilar strip of any width, as it also happens in a thick strip 
on a superconducting substrate ~\cite{cr3,gen,clem}.  

\subsection{Periodic arrays}
\begin{figure}[h!]
   	\includegraphics[scale=0.5,trim={0mm 0mm 0mm 10mm},clip]{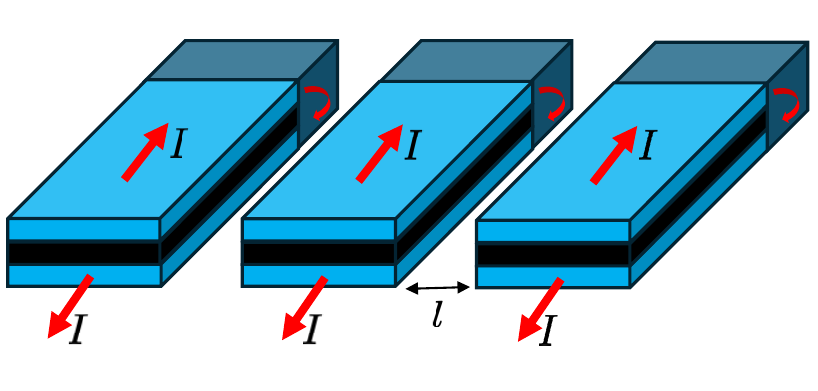}
   	\caption{Periodic array of bilayers spaced by $l$. }
   	\label{F7}
   	\end{figure}
In a periodic array of bifilar bilayers shown in Fig. \ref{F7} the neighboring bilayers are electrically disconnected and carry the same current $I$ injected by independent current supplies.    
In this case the solution for $Q(x)$ is given by Eq. (\ref{Ast}) summed up over all bilayers:
\begin{gather}
Q(x)-k\sum_{n}\int_{0}^{1}duQ(u)\times
\label{per} \\
\!\!\ln\left[\frac{[(x-u_{n})^{2}-u^{2}]^{2}}{[h^{2}+(x-u_{n}-u)^{2}][h^{2}+(x-u_{n}+u)^{2}]}\right]=\alpha,
\nonumber 
\end{gather}
where $u_n=n(2+l),\, n=0,\pm1,\pm 2,...$.
\begin{figure}[h!]
   	\includegraphics[scale=0.42,trim={0mm 0mm 0mm 0mm},clip]{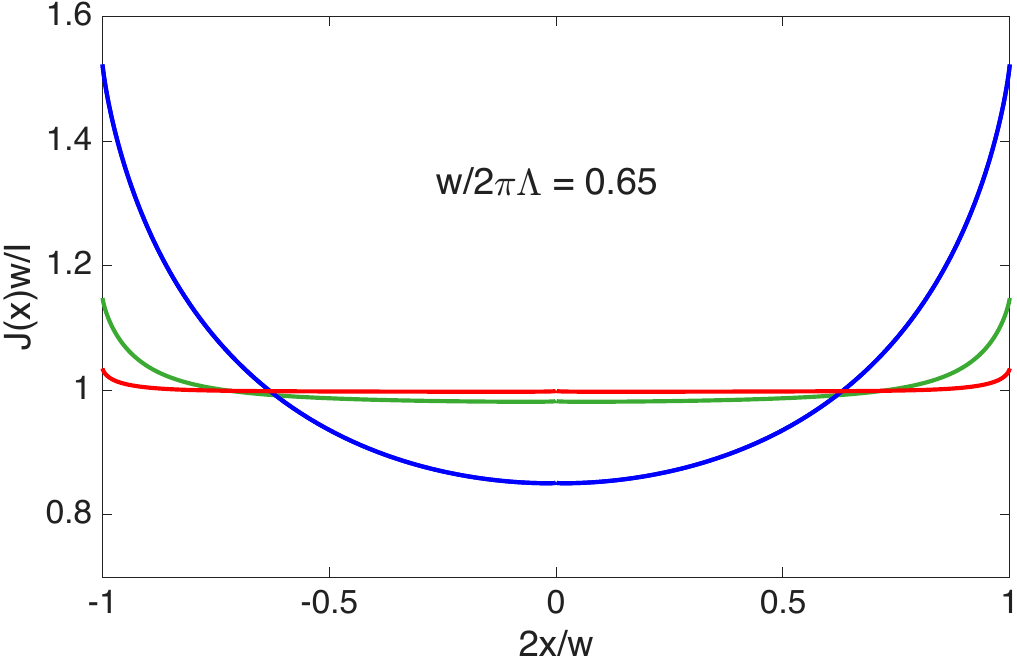}
   	\caption{Comparison of $J(x)$ for a single film (blue), single bilayer (green) and periodic array (red) calculated for $w/2\pi\Lambda=0.65$, $h=0.1w$ and $l=0.01w$.   }
   	\label{F8}
   	\end{figure}
	
Shown in Fig. \ref{F8} is $J(x)$ in the periodic array in comparison with a single bilayer and a bare strip. One can see that peaks in $J(x)$ at the edges in a periodic array are further reduced as compared to a single bilayer due to partial cancellation of magnetic fields leaking out from the neighboring bilayers. Thus, the Pearl current crowding in a periodic array of bifilar bilayers can be practically eliminated, no matter how wide they are.     

\subsection{Bilayers tuned by current}

\begin{figure}[h!]
   	\includegraphics[scale=0.4,trim={0mm 0mm 0mm 0mm},clip]{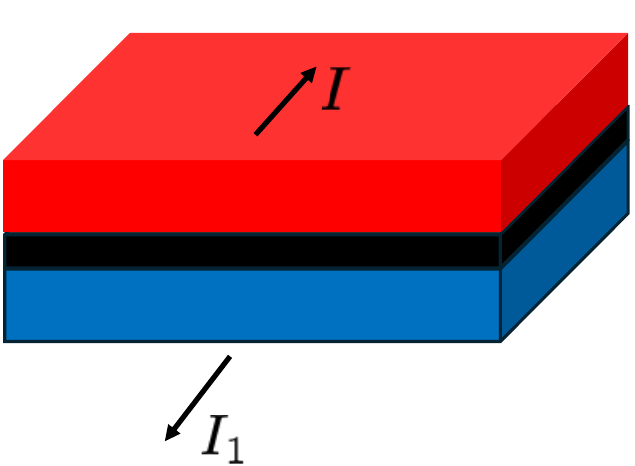}
   	\caption{A bilayer comprised of different superconducting films in which the magnetic field produced by the bottom layer is used to tune $J(x)$ in the top layer.}
   	\label{F9}
   	\end{figure}
Consider now a bilayer structure which provides no Pearl current crowding and controllable dips in $J(x)$ at the edges. This can be achieved with two disconnected strips made of different superconductors and carrying different currents $I$ and $I_1$ injected to each film separately as shown in Fig. \ref{F9}.  Here the magnetic field produced by the bottom film is used to tune $J(x)$ in the top (detector) film, similar to the side control wires discussed above. 

	\begin{figure}[h!]
   	\includegraphics[scale=0.45,trim={0mm 0mm 0mm 0mm},clip]{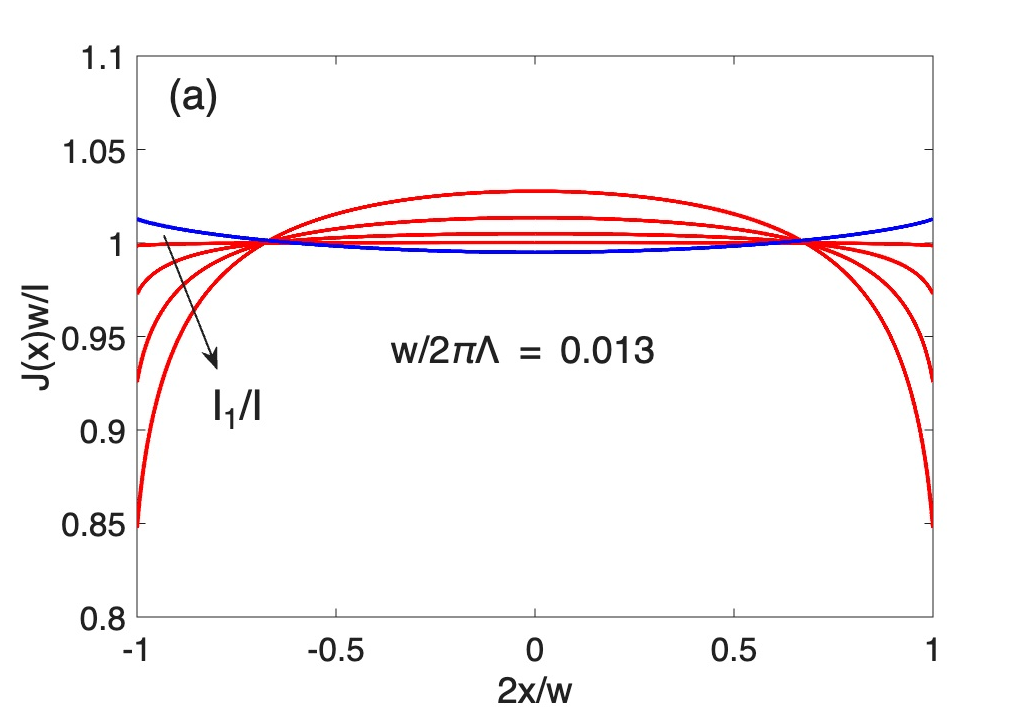}
	\includegraphics[scale=0.45,trim={0mm 0mm 0mm 0mm},clip]{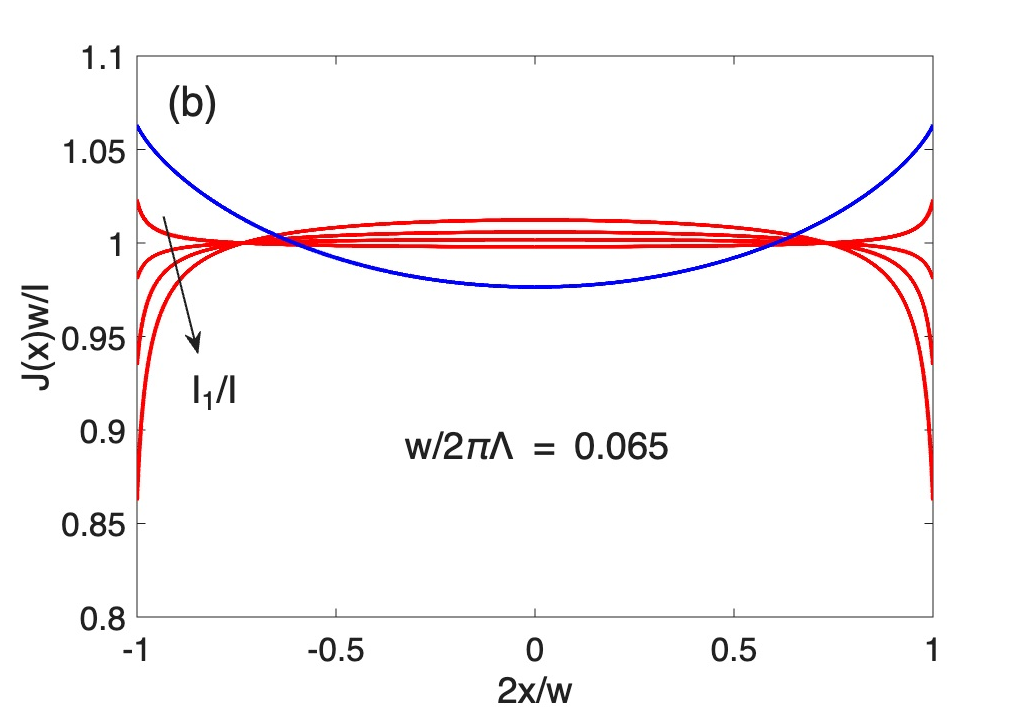}
	\includegraphics[scale=0.4,trim={0mm 0mm 0mm 0mm},clip]{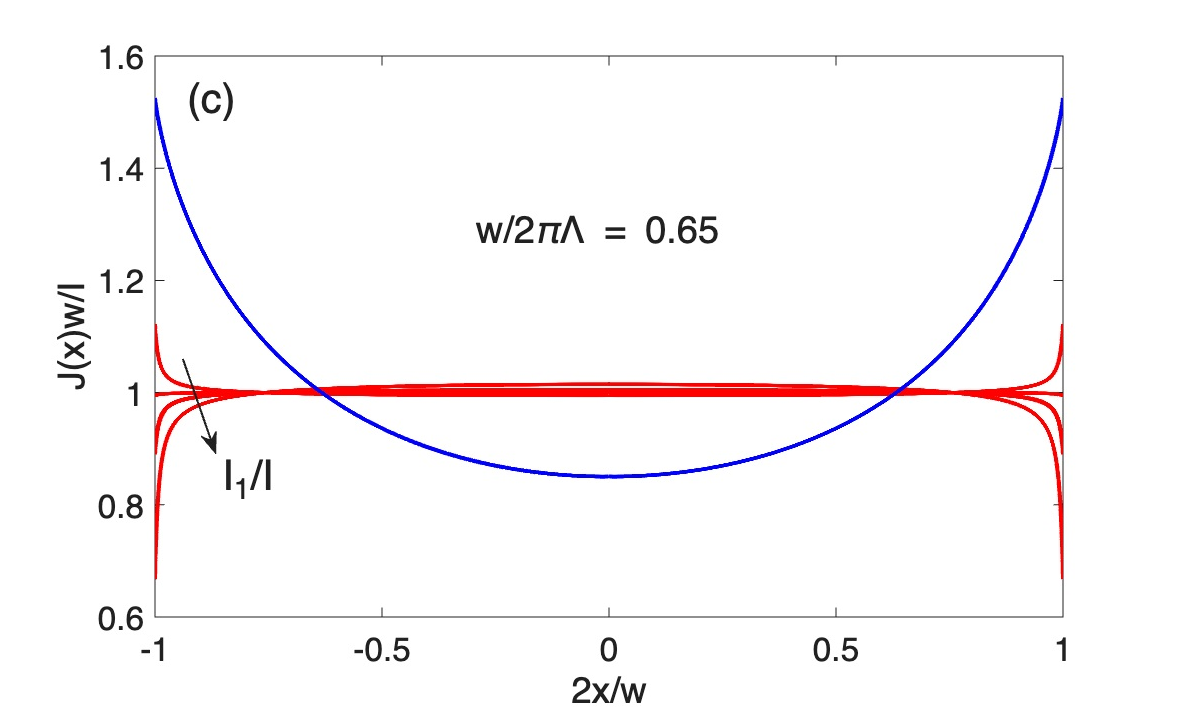}
   	\caption{$J(x)$ in the top layer of different widths at varying currents in the bottom layer, $k_1=200k$ and $h=0.0025w$:  (a) $w\approx 0.08\Lambda$, $I_1/I=-1.2, -4.61, -10.81,- 21.06$; (b) $w\approx 0.4\Lambda$, $I_1/I=0, -1.88, -3.94, -7.17$; (c) $w\approx 4\Lambda$, $I_1/I= 0, -1.09, -1.93, -3.92$. For a W$_{0.8}$Si$_{0.2}$ film with $d=4$ nm and $\Lambda = 245\,\mu$m ~\cite{wsi}, Figs. 10a-c correspond to $w\approx 20\,\mu$m, $w\approx 100\,\mu$m and $w\approx 1$ mm, respectively. The blue lines show $J(x)$ in single strips.}
	\label{F10}
	\end{figure}
	
The equations for $Q(x)$ and $Q_1(x)$ in the top and the bottom films, respectively are obtained in the same way as before for a detector film with current control wires:
	\begin{gather}
	Q(x)-2k\int_{0}^{1}\ln|x^{2}-u^{2}|Q(u)du-
	\label{bl1} \\
	\!\!\!k_1\int_{0}^{1}\ln[(u^{2}+h^{2}-x^{2})^{2}+4h^{2}x^{2}]Q_1(u)du=\alpha
	\nonumber
	\end{gather} 
	\begin{gather}
	Q_1(x)-2k_1\int_{0}^{1}\ln|x^{2}-u^{2}|Q_1(u)du-
	\label{bl2} \\
	\!\!\!k\int_{0}^{1}\ln[(u^{2}+h^{2}-x^{2})^{2}+4h^{2}x^{2}]Q(u)du=\beta
	\nonumber
	\end{gather} 
Solutions of these equations presented in Fig. \ref{F10} show that the control underlayer with $\Lambda_1\ll\Lambda$ flattens $J(x)$ in most part of the top layer, while reducing $J(x)$ at the edges. The tuning effects of side control wires and the underlayer on $J(x)$ in the strip are similar, but the bilayers may be advantageous for SNSPD arrays in multi-pixel single photon cameras as they can provide higher packing density with less wiring.

\section{Nonlinear current pairbreaking} \label{S3}
	
The above results obtained from the London theory imply that ${\bf J}=-2{\bf Q}/\mu_0\Lambda$ depends linearly on $Q$. Generally, $J(Q)$ is a nonlinear function of $Q$ at $J\sim J_d$ with $\partial J/\partial Q=0$ at $J=J_d$. This nonlinearity can be essential at the edges of single strips due to the Pearl current crowding. Yet in SNSPDs tuned by control wires the effect of nonlinearity of $J(Q)$ is mild and the inverted profiles of $J(x)$ are adequately described by the London theory in the entire range of currents up to $I_b\simeq (0.6-0.8)I_d$. 

\begin{figure}[h!]
   	\includegraphics[scale=0.45,trim={0mm 0mm 0mm 0mm},clip]{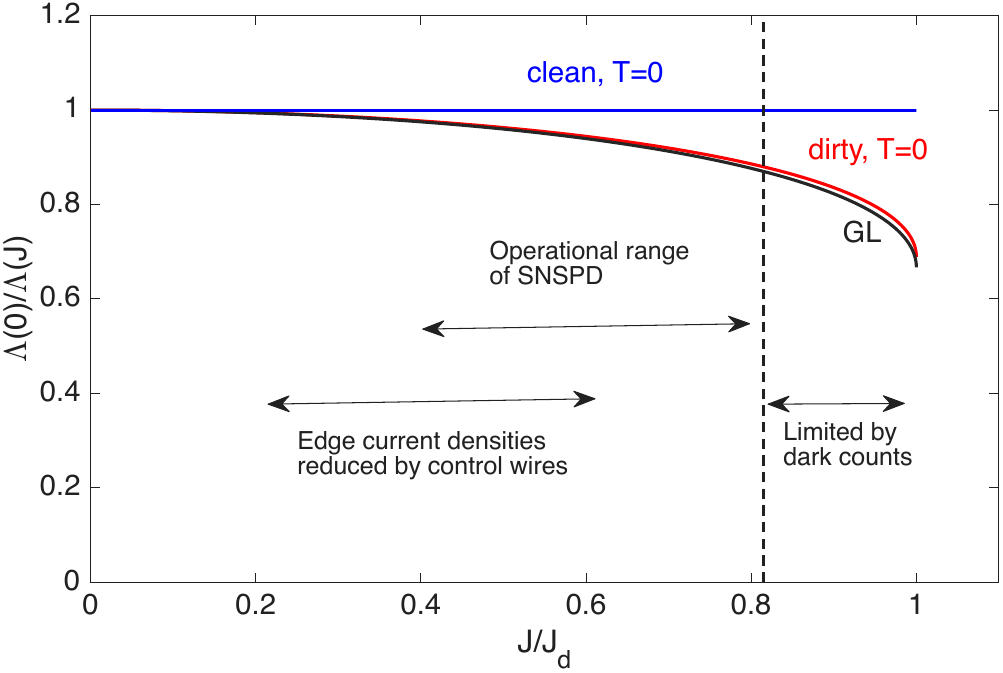}
   	\caption{$\Lambda(J)$ as functions of $J$ at $T=0$ given by the Maki theory in the clean and the dirty limits~\cite{maki} along with $\Lambda(J)$ of the GL theory. Arrows show typical operational ranges of SNSPD and reduced $J$ at the edges calculated in this work.}
	\label{F11}
	\end{figure}

The London theory is applicable if $\Lambda(J)$ depends weakly on $J$. Figure \ref{F11} shows $\Lambda (J)$ calculated from the Maki theory ~\cite{maki,ck} at $T=0$ in the clean ($l\gg \xi_0$) and the dirty ($l\ll \xi_0$) limits, along with $\Lambda (J)$ of the GL theory. 
At $0<J<0.8J_d$ the change in $\Lambda(J)$ does not exceed $10-12\%$. At $T\ll T_c$, the Pearl length $\Lambda(J)$ is somewhere between the blue and the red curves, depending on the materials purity. In the clean limit $\Lambda$ is independent of $J$ and the London theory at $T=0$ is exact up to very close to $J=J_d$~ \cite{maki}. For disordered films used in SNSPD,  $\Lambda(J)$ is a bit higher than the red line in Fig. \ref{F11}, the GL theory extrapolated to $T\ll T_c$ exaggerates the change in $\Lambda(J)$. Here $I=I_d$ cannot be reached because of unbinding of vortex-antivortex pairs, which sets the ultimate performance limit for SNSPD, as shown below. In what follows manifestations of nonlinearity of $J(Q)$ on $J(x)$ in thin film strips are addressed.

\subsection{GL current crowding in a strip}

\begin{figure}[h!]
   	\includegraphics[scale=0.42,trim={0mm 0mm 0mm 0mm},clip]{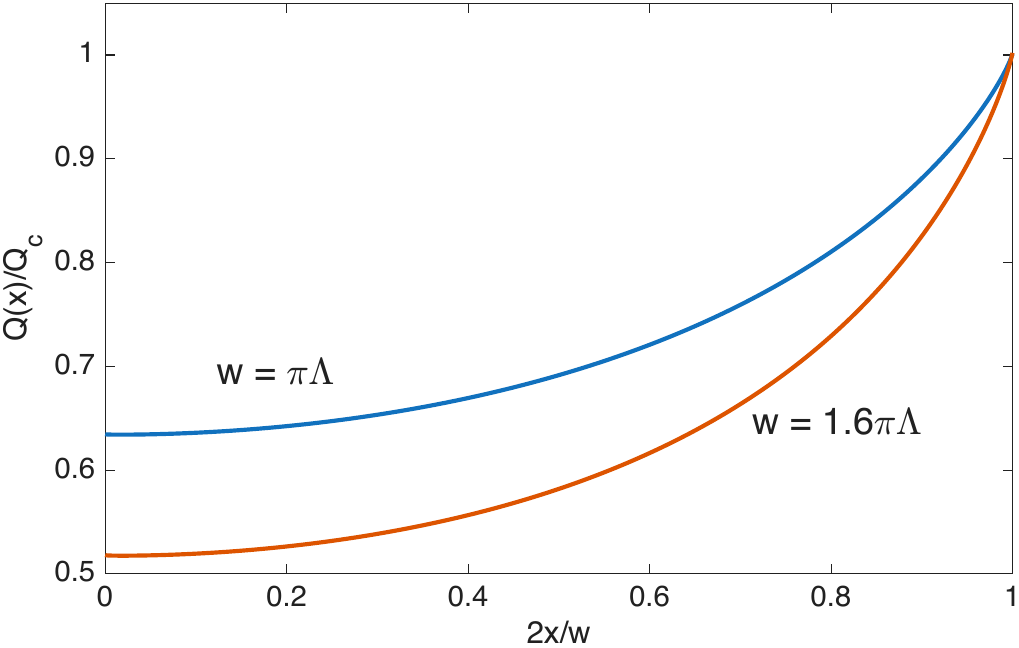}
   	\caption{$Q(x)$ calculated by solving Eq. (\ref{nino}) at currents for which $Q(x)$ at the edge reaches the GL depairing limit $Q_c=Q_{c0}/\sqrt{3}$ for $w=\pi\Lambda$ and $w=1.6\pi\Lambda$.   }
	\label{F12}
	\end{figure}

The GL theory exaggerates the effect of nonlinearity of $J(Q)$ on $\Lambda(J)$ in at $T\ll T_c$ but it is close to 
the BCS results in the dirty limit at $T=0$, as shown in Fig. \ref{F11}.  For this reason, the simpler GL theory in which 
\begin{equation}
{\bf J}=-2{\bf Q}[1-(Q/Q_{c0})^2]/\mu_0\Lambda
\label{jgl}
\end{equation}   
is used here to evaluate the effect of current pairbreaking on $J(x)$ in thin films. 
Consider first a single strip for which the equation for a dimensionless superfluid velocity $q(x)=Q(x)/Q_{c0}$ with $Q_{c0}=\phi_0/2\pi\xi$ is obtained in the same way as Eq. (\ref{Q0}):
\begin{equation}
q(x)-2k\!\int_0^1\ln|x^2-u^2|q(u)[1-q^2(u)]du=\gamma,
\label{nint}
\end{equation}
where $\gamma=\theta'\xi$ and the GL nonlinearity affects the self-field nonlocal term. For a narrow strip with 
$k\ll 1$, Eq. (\ref{nint}) yields constant $q\to\gamma$ and $J\to\phi_0\theta'[1-(\xi\theta')^2]/\pi\mu_0\Lambda$, the depairing limit is reached at $\gamma=1/\sqrt{3}$. For wider strips, Eq. (\ref{nint}) was solved iteratively: 
\begin{equation}
q_{n+1}(x)=2k\!\int_0^1\ln|x^2-u^2|[q_n(u)-q_{n}^3(u)]du+\gamma,
\label{nino}
\end{equation}
which gives rapidly converging solutions at $w\leq 5\Lambda$. Figure \ref{F12} shows $Q(x)$ calculated from Eqs. (\ref{nino}) at currents for which $Q(\pm w/2)$ at the edges reaches the GL depairing limit $Q_c=Q_{c0}/\sqrt{3}=\phi_0/2\sqrt{3}\pi\xi$.  
Here the inhomogeneity of $Q(x)$ increases with the strip width, similar to the Pearl screening in the London theory. 

\begin{figure}[h]
   	\includegraphics[scale=0.42,trim={0mm 0mm 0mm 0mm},clip]{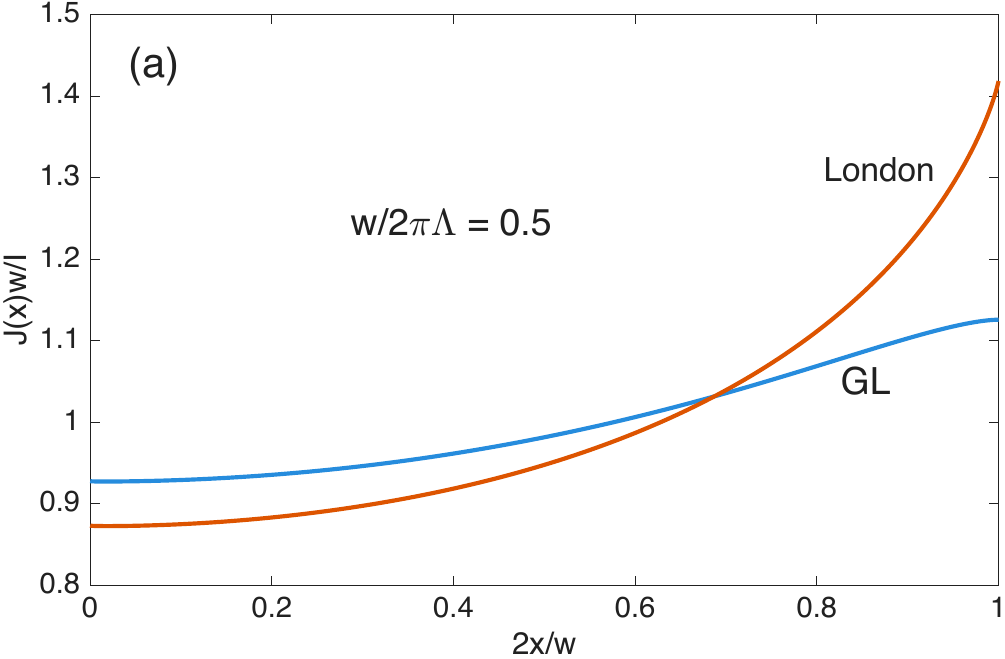}
	\includegraphics[scale=0.42,trim={0mm 0mm 0mm 0mm},clip]{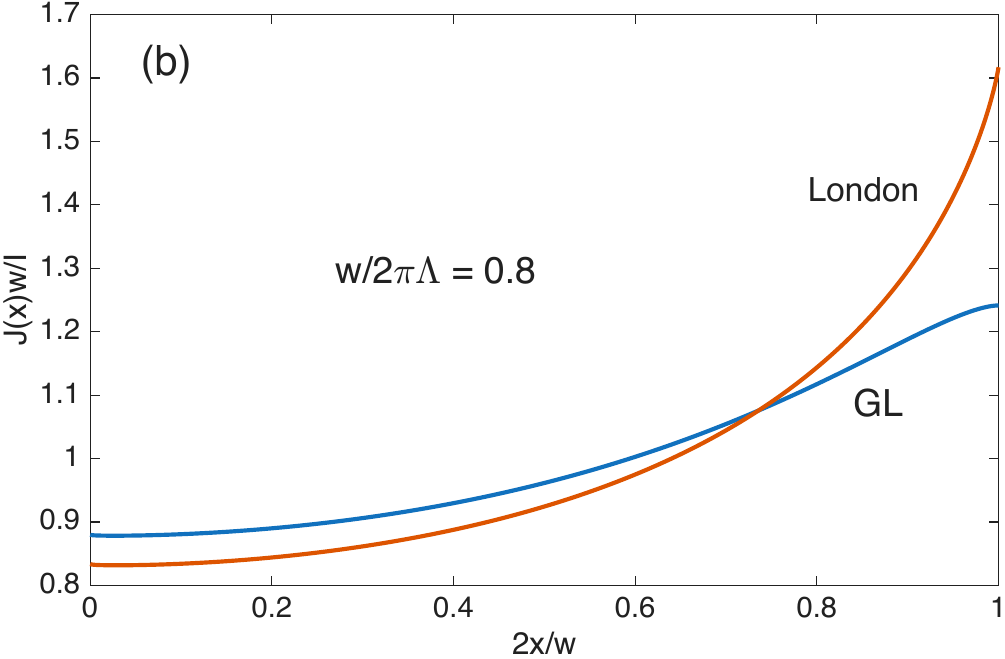}
   	\caption{$J(x)$ calculated from the London and the GL theories at the same $\gamma$ at which $J(\pm w/2)$ at the edges reaches $J_d$ in strips of  widths: (a) $w=\pi\Lambda$,  (b) $w=1.6\pi\Lambda$. }
	\label{F13}
	\end{figure}

\begin{figure}[h!]
   	\includegraphics[scale=0.42,trim={0mm 0mm 0mm 0mm},clip]{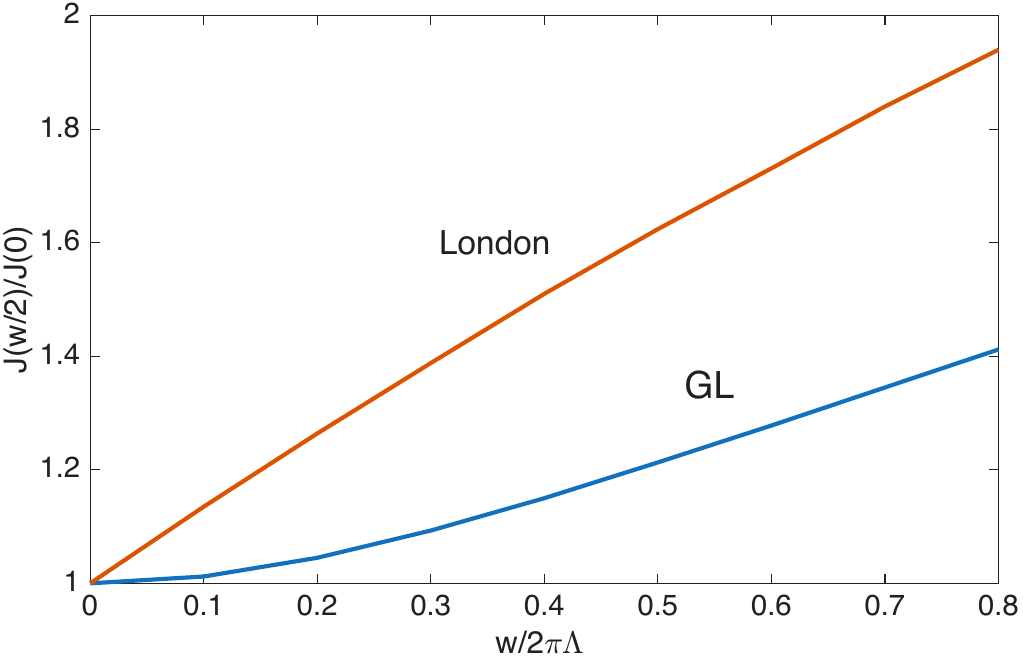}
   	\caption{The current crowding ratio $J(w/2)/J(0)$ calculated in the London limit of $I\ll I_d$ and the GL pairbreaking limit 
	for which $J(x)$ at the edge reaches $J_d$.}
	\label{F14}
	\end{figure}

The sheet current density calculated from Eqs. (\ref{jgl}) and (\ref{nino}) is less inhomogeneous than $J(x)$ in the London theory, as shown in Fig. \ref{F13}. Here the shape of $J(x)$ changes from the red one at $I\ll I_d$ to the blue one at $I\sim I_d$  so the Pearl current crowding weakens as $I$ is increased. Figure \ref{F14} shows the current crowding ratio $J(w/2)/J(0)$ as a function of the normalized strip width $w/2\pi\Lambda$ calculated from the London theory at $I\ll I_d$ and the GL theory at $I\sim I_d$. As $I$ is increased, the ratio $J(w/2)/J(0)$ for a given $w$ decreases from its maximum on the red curve at $I\ll I_d$ to its minimum on the blue curve at $I\approx I_d$, so the London model  overestimates the Pearl current crowding in wide strips at $I\sim I_d$.    

\subsection{Strip with side control wires}	
The Pearl screening amplifies the effect of the GL nonlinearity on $J(x)$ at the edges of a single strip, but the situation reverses in a strip between the   control wires which cause dips in $J(x)$ at the edges, where the effect of nonlinearity on $J(x)$ diminishes. This case is described by the following equations generalizing  Eqs. (\ref{Q0}) and (\ref{Q1}):
\begin{gather}
q(x)-2k\!\int_0^1\!\ln|x^{2}-u^{2}|[q(u)1-q^3(u)]du-
\label{q} \\
k_{1}\!\int_{1+b}^{1+l+b}\!\!\ln[(u^{2}+h^{2}-x^{2})^{2}+4h^{2}x^{2}]p(u)du=\gamma_0,
 \nonumber \\
p(x)-2k_{1}\!\int_{1+b}^{1+b+l}\!\!\ln|x^{2}-u^{2}|p(u)du-
\label{p} \\
k\!\int_0^1\!\ln[(u^{2}+h^{2}-x^{2})^{2}+4h^{2}x^{2}][q(u)-q^3(u)]du=\gamma_1,
\nonumber
\end{gather}
where $q=Q_0(u)/Q_{c0}$, $p=Q_1(u)/Q_{c0}$, the dimensionless phase gradients $\gamma_0=\xi\theta_{0}'$ and $\gamma_1=\xi\theta_1'$ are expressed in terms of currents $I$ and $I_1$ as described in Appendix A, and $\xi$ is the coherent length in the strip. For the sake of simplicity, the control wires of the same dimensions and material and carrying equal currents are considered.  

\begin{figure}[h]
   	\includegraphics[scale=0.45,trim={0mm 0mm 0mm 0mm},clip]{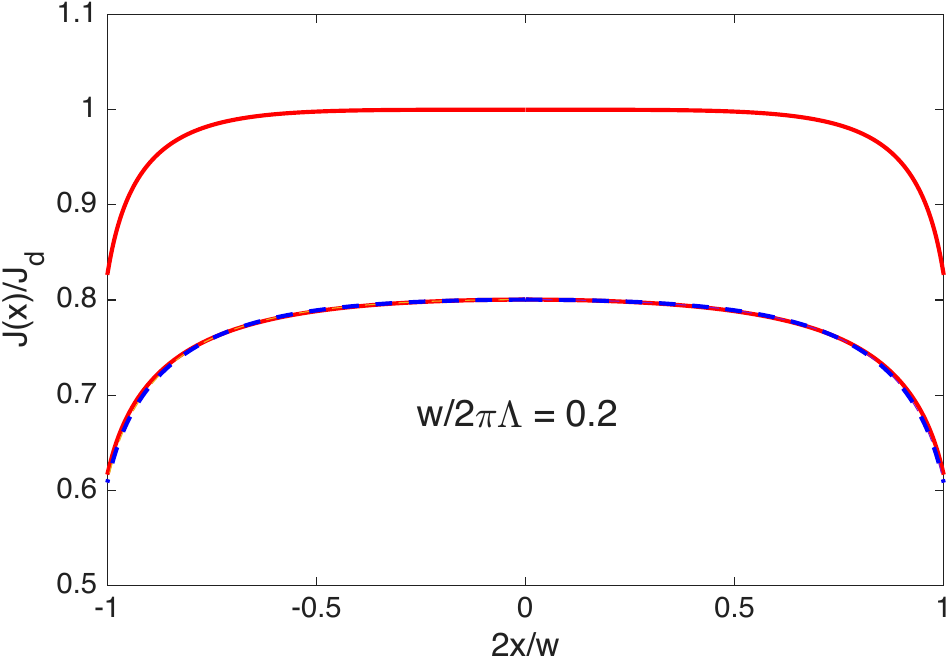}
   	\caption{$J(x)$ calculated from Eqs. (\ref{q})-(\ref{p}) at currents at which $J(x)$ reaches $J_d$  and $0.8J_d$ in the center of the strip (top and bottom red curves, respectively). The dashed blue curve is calculated in the London theory at $J(0)=0.8J_d$.  The top and bottom GL red curves correspond to $I=0.98I_d$, $I_1=1.34I$ and $I=0.77I_d$, $I_1=1.02I_d$, respectively, while the London blue curve is obtained at $I_1=0.9I$ for $k_1=500k$ and $b=h=5\cdot 10^{-3}w$. }
	\label{F15}
	\end{figure}

Equations (\ref{q}) and (\ref{p}) were solved iteratively, as described in the previous subsection, assuming the London relation $J_1=-2Q_1/\mu_0\Lambda_1$ in control wires. Shown in Fig. \ref{F15} are examples of $J(x)$ calculated from Eqs. (\ref{q}) and (\ref{p}) for the currents at which $J(x)$ in the center reaches $J_d$ and $0.8J_d$. 
The inverted profiles of $J(x)$ obtained from the London theory shown in Fig. \ref{F4} turned out to be close to those of the GL theory. For instance, Fig. \ref{F15} shows $J(x)$ calculated from the London and the GL theories at the same current $I$ at which $J(0)$ in the center reaches $0.8J_d$. Both $J(x)$ profiles coincide with the accuracy better than $1.5\%$ while their respective control currents $I_1$ differ by $12\%$. Thus, the London theory adequately describes the evolution of $J(x)$ as $I_1$ is changed at bias currents up to $I\sim I_d$. Here the exact value of variable $I_1$ is  secondary as $I_1$ can always be set to mitigate penetration of vortices from the edges, even if the sample-dependent effect of lithographic defects at the edges is unknown. This flexibility makes the London-Maxwell theory an effective tool to quickly evaluate possible advantages and disadvantages of different thin film nanostructures. 
  
\begin{figure}[h]
   	\includegraphics[scale=0.43,trim={0mm 0mm 0mm 0mm},clip]{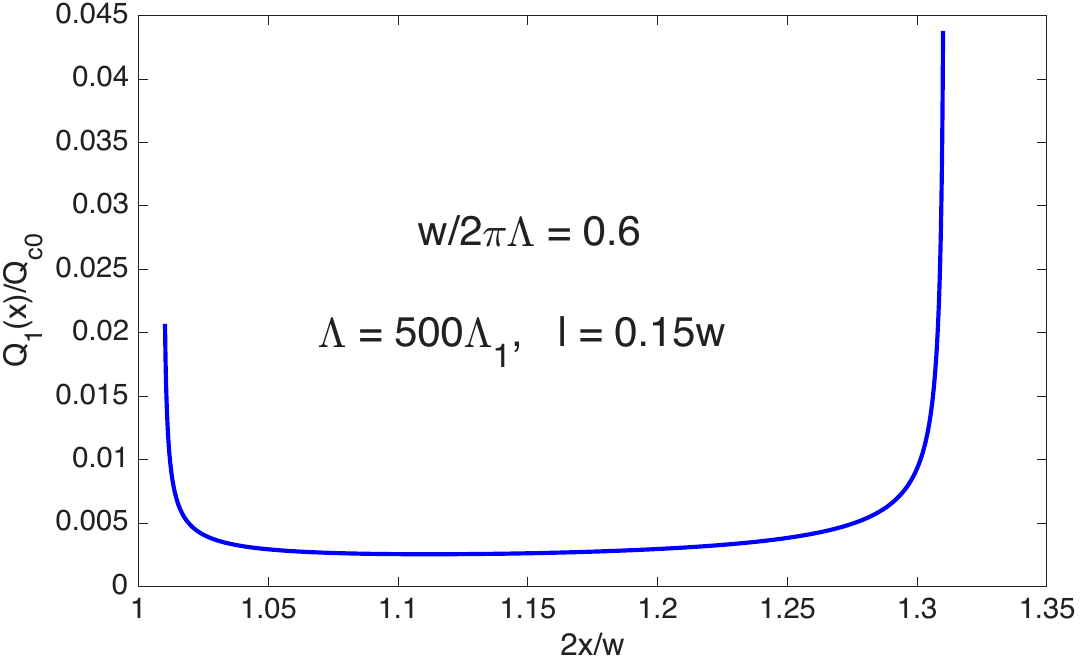}
   	\caption{$Q_1(x)/Q_{c0}$ in the control strip calculated from Eqs. (\ref{q})-(\ref{p}) at $k_1=500k$, $l=0.15w$, $b=h=5\cdot 10^{-3}w$, $w=3.77\Lambda$, $I=0.988I_d$, $I_1=0.78I$. }
	\label{F16}
	\end{figure}

The above results were obtained assuming that $J_1(x)$ in the side wires is below the pairbreaking limit. This is usually the case if $\Lambda_1\ll\Lambda$, as it was pointed out above. Shown in Fig. \ref{F16} is an example of calculated $Q_1(x)/Q_{c0}$ in the control wire in which the asymmetry of $Q_1(x)$ results from its inductive coupling with the central strip and another wire. Yet even the highest peak in $Q_1(x)\simeq 0.045Q_{c0}\simeq 0.36Q_{c1}$ is well below the GL depairing limit $Q_{c1}=\phi_0/2\sqrt{3}\pi\xi_1$ so the London electrodynamics in the side strip is applicable.   
	
Manifestations of the GL pairbreaking in the bilayer shown in Fig. \ref{F9} are similar to those in a strip between two side wires. In the bifilar structures $J(x)$ is practically flat at any width, so the account of the GL pairbreaking in Eqs. (\ref{Q2}) and (\ref{per}) reduces to solving the cubic equation $I=(wQ/2\mu_0\Lambda)[1-(Q/Q_{c0})^2]$ for $Q$ at $I<I_b$. 

 \subsection{Superconducting diode}
 Current flow in a strip coupled with control wires depends on the relative orientation of $I$ and $I_1$. The dips in $J(x)$ at the strip edges occur if currents in the strip and the wires flow in the same direction. For antiparallel $I$ and $I_1$, the self field of the wires enhances the current crowding at the edges of the strip. In a bilayer shown in Fig \ref{F9}, the current crowding is reduced if $I$ and $I_1$ are antiparallel and enhanced if both $I$ and $I_1$ flow in the same direction.   Because of such current non-reciprocality, these structures behave as superconducting diodes in which the critical currents of switching to a resistive state depend on the polarity of $I$ at a fixed $I_1$. The polarity-dependent critical currents in thick superconducting films coupled with current control wires were first observed in the sixties ~\cite{cr0,cr1,cr2}. Recently TDGL simulations of the vortex diode for the geometry shown in Fig. \ref{F2} were done in Ref. \onlinecite{didi} in which ${\bf J}({\bf r})$ was not calculated self-consistently by matching ${\bf Q}(x,y)$ to ${\bf H}(x,y)$ outside the strip but evaluated using an ad-hoc boundary condition at the strip edge incompatible with Eq. (\ref{lond}).  Here  the diode effect is calculated for a thin strip in the Pearl limit with $d\ll \lambda$ and $\Lambda\gg\lambda$ characteristic of SNSPDs with $\kappa\sim 10^2$ and $d<\xi$ ~\cite{spd1,spd2,spd3}. 
     
 \begin{figure}[h]
   	\includegraphics[scale=0.43,trim={0mm 0mm 0mm 0mm},clip]{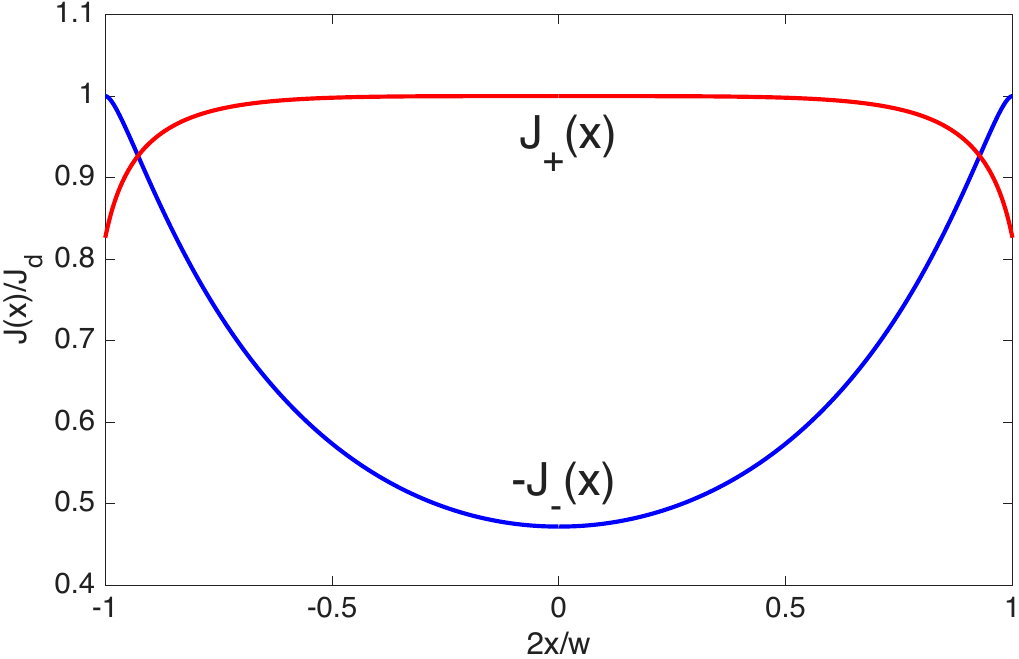}
   	\caption{Sheet current densities in a strip calculated from Eqs. (\ref{q}) and (\ref{p}) for parallel $J_+(x)$ and antiparallel $J_-(x)$ current flow with respect to a fixed control current $I_1$. The critical currents are $I_c^+=0.983I_d$ and $I_c^-=-0.63I_d$ at $I_1=1.32I_d$, $w=1.26\Lambda$, $k_1=500k$, $l=0.15w$, $b=h=5\cdot 10^{-3}w$, where $J_d$ is the depairing current density in the strip. }
	\label{F17}
	\end{figure}

Shown in Fig. \ref{F17} are the sheet current densities $J_+(x)$ and $J_-(x)$ in a strip calculated from Eqs. (\ref{q}) and (\ref{p}) for parallel and antiparallel $I$ and $I_1$, respectively. Here $I_1$ was fixed and $J_+(x)$ and $J_-(x)$ were calculated at critical currents $I_c^+$ and $I_c^-$ at which $J(x)$ reached $J_d$ either in the center or the edges of the strip.  For parallel $I$ and $I_1$, the depairing limit is first reached in the center, while for antiparallel $I$ and $I_1$, it happens at the edges.  As a result, $I_c^+$ and $I_c^-$ given by the areas under the curves in Fig. \ref{F17} become markedly different, the difference between $I_c^+$ and $I_c^-$ increases as $I_1$ is increased.
 
\begin{figure}[h]
   	\includegraphics[scale=0.43,trim={0mm 0mm 0mm 0mm},clip]{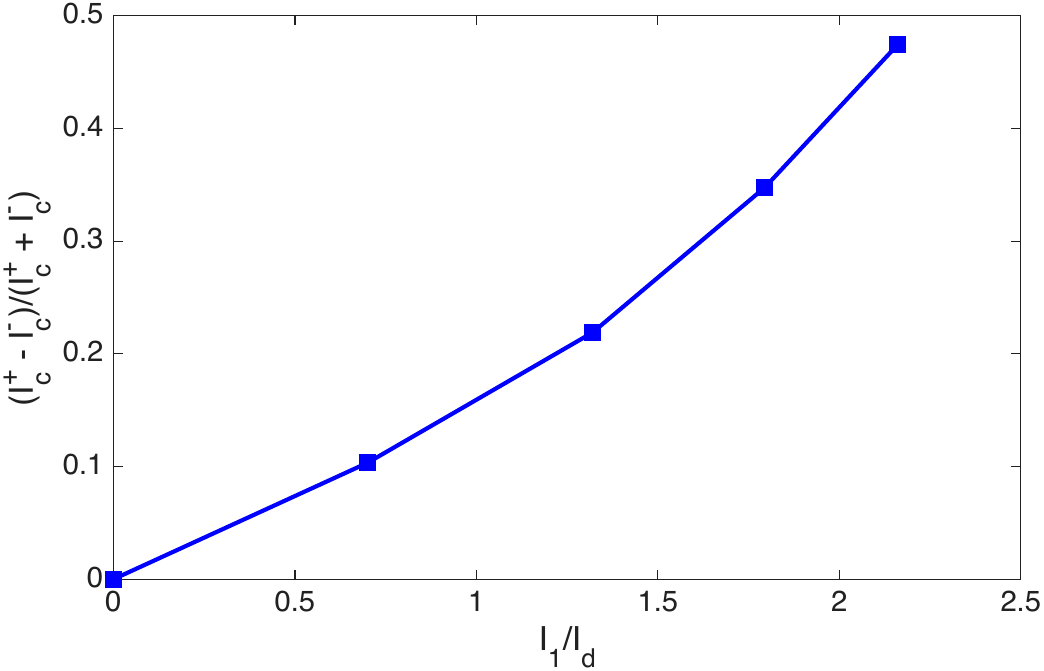}
   	\caption{The diode efficiency parameter as a function of the normalized control current calculated at $w=1.26\Lambda$, $k_1=500k$, $l=0.15w$, $b=h=5\cdot 10^{-3}w$. }
	\label{F18}
	\end{figure}
	
The critical currents $I_c^+$ and $I_c^-$ were calculated for different $I_1$ ranging from $0$ to $2.2I_d$ and other parameters specified in the caption of Fig. \ref{F17}. Using the so-obtained $I_c^+(I_1)$ and $I_c^-(I_1)$, the diode efficiency $(I_c^+ - I_c^-)/(I_c^++I_c^-)$ was calculated as a function of $I_1$. The result shown in Fig. \ref{F18} suggests that such vortex diode controlled by current {\it in-situ} can be quite efficient. This distinguishes this superconducting diode from other propositions in which the non-reciprocal current response requires an external magnetic field combined with spatial inhomogeneity of materials properties, edge indentations, superconducting-ferromagnetic structures, spin-orbital effects, or unconventional superconductivity with nonzero momentum of Cooper pairs ~\cite{di1,di2,di3}.    

\section{Vortex resistive states tuned by control wires. }
\label{secA5}

\begin{figure}[h]
\centering
   	\includegraphics[scale=0.5,trim={5mm 0mm 0mm 0mm},clip]{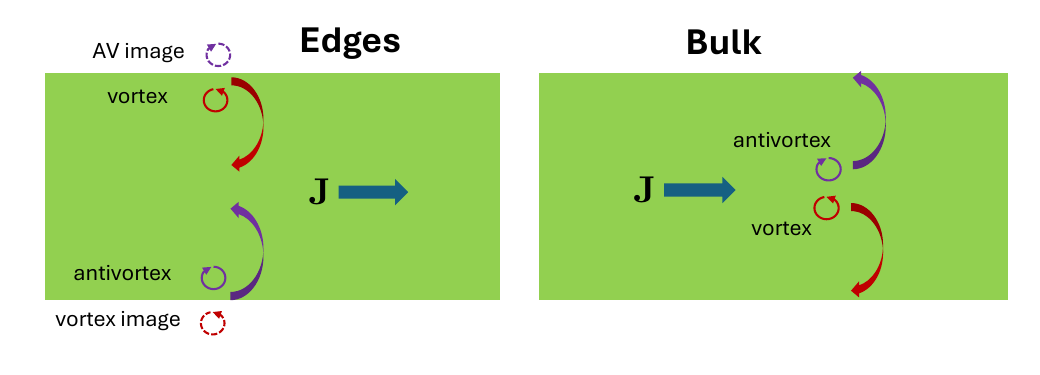}
   	\caption{Single vortices and antivortices penetrating from the film edges and VAV pair unbinding inside the strip.}
   	\label{F19}
   	\end{figure}

Engineering supercurrent flow in a strip by control wires allows tuning resistive states caused by the motion of vortices. As $I$ exceeds either $I_c^+$ or $I_c^-$, the strip switches to a flux flow state. For parallel $I$ and $I_1$, there are two possibilities: 1. $I_1$ is not strong enough to deactivate the Pearl current crowding and the edge defects. In this case the resistance is caused by penetration of vortices and antivortices from the opposite edges of the strip. 2. If $I_1$ is sufficient to deactivate the current crowding and the edge defects, the resistive transition occurs due to unbinding of vortex-antivortex (VAV) pairs ~\cite{bkt1,bkt2}. For antiparallel $I$ and $I_1$, the enhanced current crowding at the edges ensures that the resistance at $I>I_c^-$ is caused by penetration of vortices and antivortices and their annihilation in the center, as depicted in Fig. \ref{F19}

This section focuses not on a self-field flux flow resistance at $I>I_c^{\pm}$ affected by the control wires ~\cite{didi,vt1,vt2} but on a small voltage $V(I,T)$ at subcritical currents. The behavior of $V(I,T)$ at $I<I_c^\pm$ determines the dark count rate in photon detectors and also can be a sensitive indicator of the transition from the edge-dominated to the bulk-dominated resistance. This transition occurs above the threshold control current $I_1>I_1^*$ which depends on the strip width and the current crowding effect of edge defects. Here the thermally-activated uncorrelated hopping of sparse vortices and non-overlapping VAV pairs is considered, leaving aside multi-vortex effects near the BKT transition at higher currents ~\cite{bkt2} and quantum tunneling of vortices at ultra low $T$ ~\cite{tun1,tun2}.    

The rate $S_e$ of thermally-activated hopping of vortices through the edge is determined by the energy of the vortex $E(u)$ which can be written as a work to move a vortex by a distance $u$ from the edge of the strip:
\begin{equation}
E(u)=E_v(u)-\phi_0\int_0^uJ(x)dx,
\label{euu}
\end{equation}
where $E(u) =0$ at the strip edges ~\cite{gurvink,stejic,vod}. The last term in Eq. (\ref{euu}) is the work of the Lorentz force of transport current density $J(x)$ calculated above for different geometries. The self-energy $E_v(u)$ is the work against the attraction force between the vortex and the edges. For a strip of arbitrary width, $E_v(u)$ evaluated by the method of current images ~\cite{gurvink} was derived in Appendix B:
\begin{gather}
E_{v}(u)=\epsilon\ln\left[\frac{w}{\pi\tilde{\xi}}\sin\left(\frac{\pi u}{w}\right)\right]+
2\epsilon\sum_{n=1}^\infty\sin^2\left(\frac{\pi un}{w}\right)\times 
\nonumber\\
\left[\frac{4}{\pi\sqrt{n^2-(w/\pi\Lambda)^2}}\tan^{-1}\left[\frac{n-w/\pi\Lambda}{n+w/\pi\Lambda}\right]-\frac{1}{n}\right].
\label{uoo}
\end{gather}
Here $\epsilon=\phi_{0}^{2}/2\pi\mu_{0}\Lambda$ is the vortex line energy,  $\tilde{\xi}=0.34\xi$ and the factor 0.34 accounts for the vortex core energy obtained from the GL calculations \cite{stejic,vod}.  For a narrow strip with $w\ll \pi\Lambda$, Eqs. (\ref{euu}) and (\ref{uoo}) yield ~\cite{gurvink,vod}:
\begin{equation}
E(u)=\epsilon\ln\left(\frac{w}{\pi\tilde{\xi}}\sin\frac{\pi u}{w}\right)-\phi_{0}Ju.
\label{E1}
\end{equation}
This result (without the core energy) first obtained in  Ref. \onlinecite{kkl} also follows from a  complex velocity potential of a vortex in an ideal liquid in a channel~ \cite{tfkp}.     

Equation (\ref{uoo}) does not fully account for nonlocal coupling of vortex currents via the long-range magnetic field outside the strip, which results in a complex equations for $Q(x)$ ~\cite{kogan}. Yet Eq. (\ref{uoo}) gives exact asymptotic of $E_v(u)$ at $u<\Lambda$, where the main contribution comes from the unscreened interaction of the vortex with the closest AV image ~\cite{vgk} and $u\gg \Lambda$, where it reproduces the energy of the Pearl vortex in an infinite film.  At $w\sim \Lambda$ Eq. (\ref{uoo}) interpolates between these two limits.     
\begin{figure}[h]
\centering
   	\includegraphics[scale=0.5,trim={0mm 0mm 0mm 0mm},clip]{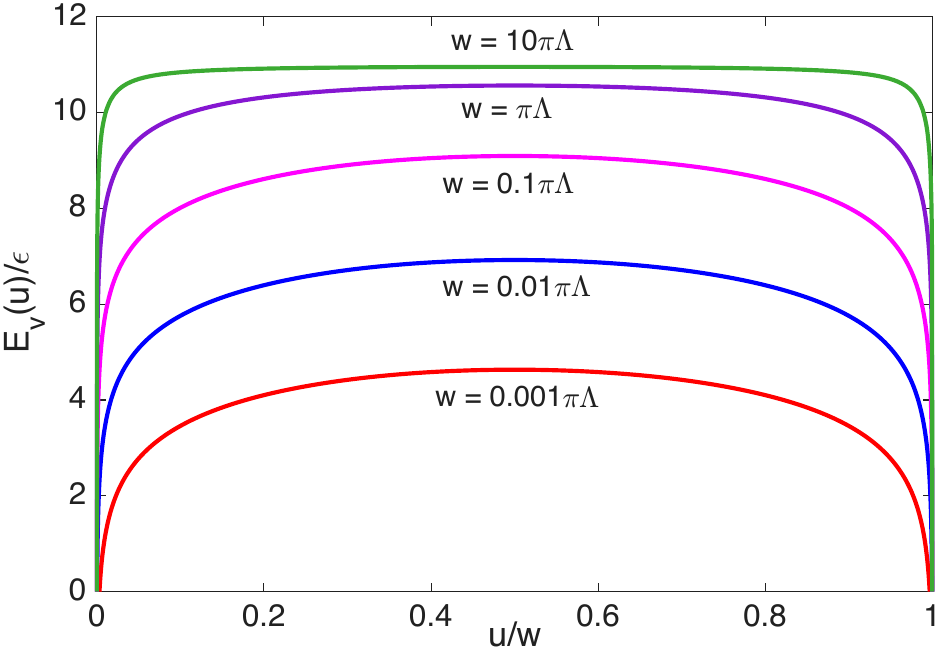}
   	\caption{The vortex self energy $E_{v}(u)$ calculated from Eq. (\ref{uoo}) for different strip widths with the parameters $\Lambda=245\,\mu$m, 
	$\xi=7$ nm of a 4 nm thick W$_{0.8}$Si$_{0.2}$ film ~\cite{wsi}. Narrow regions of $u\lesssim \tilde{\xi}$ at the edges, where the London theory is not applicable are excluded.}
   	\label{F20}
   	\end{figure}

Shown in Fig. \ref{F20} is $E_{v}(u)$ calculated from Eq. (\ref{uoo}). As $w$ is increased, $E_v(u)$ flattens in the  central part of the strip and approaches $E_v\simeq \epsilon\ln(\Lambda/\xi)$ at $w\gg \Lambda$ (see also Ref. \onlinecite{kogan}). The energy barrier of vortex hopping across the strip $U\simeq \epsilon\ln (w/\tilde{\xi})$ at $J=0$ increases slowly with the width at $w< \Lambda$ and levels off at $\simeq \epsilon\ln(\Lambda/\tilde{\xi})$ if $w\gg\Lambda$.  Transport current shifts the maximum in $E(u)$ from the middle of the strip to the position $u_m(J)$ closer to the edge, 
where $u_m$ readily follows from Eq. (\ref{E1}): $u_m=(w/\pi)\cot^{-1}(J/J_0)$ and $J_{0}=\pi\epsilon/\phi_{0}\sim (\xi/w)J_{d}\ll J_{d}$.

\begin{figure}[h]
\centering
   	\includegraphics[scale=0.4,trim={0mm 0mm 0mm 0mm},clip]{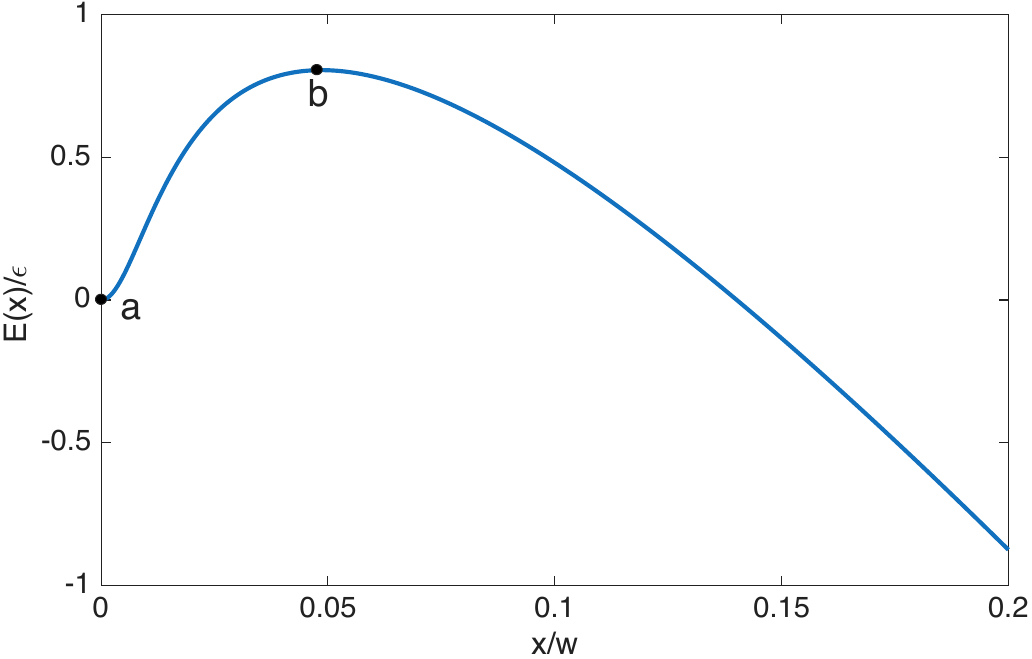}
   	\caption{Qualitative dependence of $E(x)$ on the vortex position evaluated from Eqs. (\ref{E1}) regularized by $x\to\tilde{x}= [x^2+\tilde{\xi}^2]^{1/2}$, $E(x)\to E(\tilde{x})+\phi_0J\tilde{\xi}$, $\tilde{\xi}=10^{-2}w$ and $J=20J_0$. }
   	\label{F21}
   	\end{figure}

The London model becomes invalid for a vortex spaced by $u\sim \xi $ from the edges. Numerical GL calculations of $E_v(u)$ ~\cite{vod} have shown that the London model captures the behavior of $E(u_m)$ if $J$ is not too close to $J_d$, and that the energy barrier at $J=J_d$ vanishes.  To take this effect into account, we define the energy barrier for the vortex entry as $U_e=E[u_m(J)]-E[u_m(J_d)]$. At $J\gg J_0$ this yields $U_e$ independent of the film width:
\begin{equation}
U_e=\epsilon\ln (J_d/J_e),\qquad J_e\gg J_d\xi/w.
\label{E3}
\end{equation}
This gives a reasonable approximation of $U_e(J_e)$ at $J_e\sim J_d$ at which $u_m\sim \xi\ll w$ is close to the edge of the film, as depicted in Fig. \ref{F21}. Here the current density at the strip edges $J_e(I,I_1)$ was calculated above for different geometries.  Thermally-activated hopping of vortices over the edge barrier results in a mean dc voltage     
$V =V_0(J_e/J_d)^{\epsilon/T}$, where $V_0$ was evaluated in Ref. \onlinecite{gurvink}.  At $\epsilon/T\gg1$ and $J_e\simeq J_{d}$ the V-I curve  takes the form: 
\begin{equation}
V=V_0\exp[-U_e/T],\qquad U\simeq\epsilon(1-J_e/J_d).
\label{E4}
\end{equation}

The probability of thermally-activated penetration of a vortex from the film edge is proportional to the Arrhenius factor $\propto\exp(-U_e/T)$ which can increase substantially due to edge defects which locally reduce $U[z,J(z)]$ and facilitate penetration of vortices, as shown in Fig. \ref{F1}. Here the averaged Arrhenius factor $\langle \exp[-U_e(J,z)/T\rangle$ is dominated by strong defects like sharp indentations causing significant local current crowding ~\cite{def1,def2}. Given many uncertainties depending on the film growth and deposition parameters and also on whether the edge cross-section is rectangular or rounded ~\cite{edge,geomb}, it is assumed here that edge defects locally increase an averaged current density in a narrow belt along the edge depicted by a dashed line in Fig. \ref{F1}. In this case $U_e(I)=\epsilon(1-\zeta_1J_e/J_d)$, where the current crowding factor $\zeta_1>1$ accounts for the defect-mediated enhancement of $J(x)$ at the edges.
 
The number $S_e$ of thermally-activated vortices penetrating over the edge barrier per unit time in a strip of length $L$ can be evaluated as follows:
\begin{equation}
S_e = S_{e0}\exp\biggl[-\frac{\epsilon}{T}\left(1-\frac{\zeta_1J_e}{J_d}\right)\biggr],\qquad S_{e0}\sim \frac{L\nu}{l_i}
\label{E5} 
\end{equation}
Here $L/\l_i$ is a number of statistically-independent vortex entries through the edge defects with a mean spacing $l_i$ and $\nu$ is an attempt frequency. Equal contributions of vortices and antivortices penetrating from the opposite edges of the strip are included in the definition of $l_i$. For overdamped Abrikosov vortices, $\nu\simeq \sqrt{k_ak_b}/4\pi\eta$ follows from the classical result of Kramers~ \cite{hanggi}, where $k_a$ and $k_b$ are curvatures of $E(x)$ at the bottom and the top of the potential well depicted in Fig. \ref{F20}, $\eta=\phi_0^2d/2\pi\xi^{2}\rho_n$ is the Mattis-Bardeen vortex drag coefficient, and $\rho_n$ is the normal state resistivity.  Taking $\sqrt{k_{a}k_{b}}\sim \phi_{0}^{2}d/4\pi\mu_{0}\lambda^2\xi^{2}$ at $J\sim \tilde{J}_d$ in Eq. (\ref{E5}) yields
\begin{equation}
S_{e0}\sim \frac{LR_\square}{4\pi\mu_0\Lambda l_i}
\label{E6}
\end{equation}
For a 4 nm thick W$_{0.8}$Si$_{0.2}$ film with $L=1$ mm, $\Lambda=245\,\mu$m, $R_\square =\rho_n/d=340\, \Omega$ ~\cite{wsi} and $l_i\sim 10^2$ nm, we get $\epsilon\approx 150$ K and $S_{e0}\sim 10^{14}$ s$^{-1}$. Here Eq. (\ref{E6}) is no more than a rough estimate which takes into account neither microstructural details of edge imperfections nor deformation of the vortex core at the edge in the presence of current ~\cite{vod} nor local variations of $U_e(z,I)$ along the edges which may add a power-law factor $(1-J_e/J_d)^n$ with $n\sim 1$. Yet the logarithmic slope 
\begin{equation}
\frac{d\ln S_{e}}{dI}=\frac{\epsilon\zeta_1}{TI_d} 
\label{E7}
\end{equation}
is practically insensitive to the uncertainties in $S_{e0}$ and its possible power-law dependence on $I$ if $T\ll\epsilon$.  
From Eq. (\ref{E5}), one can get a maximum current $I_{c1}$ which can flow in SNSPD at an operational dark count rate $S_c$:
\begin{equation}
I_{c1}=\frac{I_d}{\zeta_1\zeta_2}\left(1-\frac{T}{\epsilon}\ln\frac{S_{e0}}{S_c}\right),
\label{E8}
\end{equation} 
where $\zeta_2$ relates $J_e=\zeta_2\bar{J}$ with the mean current density $\bar{J}=I/w$. The factor $\zeta_2$ can be larger than 1 in a bare strip or smaller than 1 in a strip with control wires producing an inverted $J(x)$ profile. For $S_c=10^2$ s$^{-1}$, $S_{e0}=10^{14}$ s$^{-1}$ and $T/\epsilon = 1/150$ at 1 K, Eq. (\ref{E8}) yields $I_{c1}\approx 0.82 I_d/\zeta_1\zeta_2$. Here $I_{c1}$ decreases logarithmically with $L$. The account of a factor $(1-I/\tilde{I}_d)^n$ in $S_{e0}$ only increases $I_{c1}$ by few $\%$ and is neglected hereafter. 

Another contribution to $S$  comes from thermally-activated unbinding of VAV pairs of radius $r_p\simeq 2\xi J_d/J(x)\ll w$ in the strip ~\cite{bkt1}. The edge and the bulk contributions to $S$ are illustrated in Fig. \ref{F19}, from which it follows that the energy scale of the VAV pair unbinding inside a strip with $w\gg\xi$ at $J\sim J_d$ is twice that of the vortices penetrating from the edges: 
\begin{equation}
U_b(x)=2\epsilon\left[1-\frac{J(x)}{J_d}\right],\qquad J(x)\sim J_d
\label{E9}
\end{equation}
where $U_b(x)$ varies slowly over the VAV pair radius $r_p\simeq 2\xi J_d/J(x)\ll w$. Neither Eq. (\ref{E3}) for an ideal edge nor Eq. (\ref{E9}) for a uniform strip are applicable as $J$ gets close to $J_d$ at which the description in terms of uncorrelated vortices and non-overlapping VAV pairs breaks down ~ \cite{bkt1,bkt2}. Here the notion of the energy barrier for a single vortex becomes ill-defined as the uniform current-carrying state at $J=J_d$ is unstable with respect to infinitesimal perturbations of the order parameter along the entire strip.  Yet if $I_{c1}$ and $I_{c2}$ are not too close to $I_d$, the linear approximations of $U_e(J)$ and $U_b(J)$ given by Eqs. (\ref{E4}) and (\ref{E9}) are applicable.    
   
The production rate of uncorrelated VAV pairs in a film at $I\sim I_d$ is evaluated  as follows:
\begin{equation}
S_b=\nu L\int_0^w\exp\left[-\frac{2\epsilon}{T}\left(1-\frac{J(x)}{J_d}\right)\right]\frac{dx}{2\pi\xi^2}.
\label{E10}
\end{equation}
Here $A/2\pi\xi^2$ is a number of statistically-independent positions of the vortex core of radius $\sqrt{2}\xi$ in a film of area $A=Lw$.
The total $S$ is a sum of $S_e$ and $S_b$:
\begin{eqnarray}
S=\nu L\bigg[\frac{1}{l_i}e^{-\frac{\epsilon}{T}\left(1-\frac{\zeta_1J_e}{J_d}\right)}+
\label{S} \\
\int_{-w/2}^{w/2}e^{-\frac{2\epsilon}{T}\left(1-\frac{J(x)}{J_d}\right)}\frac{dx}{2\pi\xi^2}\bigg].
\nonumber
\end{eqnarray}
This equation expresses the dark count rate $S(I,I_1)$ in terms of $J(x)$ and $J_e=J(\pm w/2)$ calculated above for different  geometries. In a narrow strip with no control wires $J(x)$ is uniform so the edge contribution enhanced by the defect parameter $\zeta_1>1$ dominates $S(I,T)$ at $T\ll \epsilon$. The edge contribution is further enhanced in wide strips with  $w>\Lambda$ due to the Pearl current crowding.

The control wires can reduce $J_e$ to the point at which $S_e$ becomes smaller than $S_b$.  The inverted $J(x)$ profiles have a weak maximum at the center, where $J(x)\approx J(0)-x^2J''/2$. If $w^2\epsilon J''\gtrsim 4T$ the VAV contribution mostly comes from a central part of the strip.  In this case the integrand in Eq. (\ref{S}) is peaked at $x=0$ so the integration limits can be extended to infinity, giving:
\begin{eqnarray}
\!\!S=S_{e0}\bigg[e^{-\frac{\epsilon}{T}\left(1-\frac{\zeta_1J_e}{J_d}\right)}+
\frac{l_i}{2\xi^2}\sqrt{\frac{TJ_d}{\pi\epsilon J''}}e^{-\frac{2\epsilon}{T}\left(1-\frac{J(0)}{J_d}\right)}\bigg].
\label{S0}
\end{eqnarray}
If $J(x)$ is sufficiently flat in the most part of strip $(w^2\epsilon J''\lesssim 4TJ_d)$, Eq. (\ref{S}) becomes:
 \begin{eqnarray}
S=S_{e0}\left[e^{-\frac{\epsilon}{T}\left(1-\frac{\zeta_1J_e}{J_d}\right)}+
\frac{wl_i}{2\pi\xi^2}e^{-\frac{2\epsilon}{T}\left(1-\frac{J(0)}{J_d}\right)}\right].
\label{S00}
\end{eqnarray}
Consider for the sake of clarity the London model in which both $J_e=\zeta_2\bar{J}$ and $J(0)=\zeta_3\bar{J}$ are proportional to the mean current density $\bar{J}=I/w$ and the factors $\zeta_2<1$ and $\zeta_3>1$ depend only on $w/\Lambda$ and $I_1$. In the case described by Eq. (\ref{S00}) the VAV unbinding dominates over the edge contribution $S_e$  if:  
\begin{equation}
\frac{\bar{J}}{J_d}(2\zeta_3-\zeta_1\zeta_2)>1-\frac{T}{\epsilon}\ln\left(\frac{wl_i}{2\pi\xi^2}\right).
\label{S1}
\end{equation}
This condition is satisfied at $\bar{J}>J_v$, where
\begin{equation}  
J_v=\frac{J_d}{(2\zeta_3-\zeta_1\zeta_2)}\left[1-\frac{T}{\epsilon}\ln\left(\frac{wl_i}{2\pi\xi^2}\right)\right].
\label{Jv}
\end{equation}
For $l_i=10\xi$, $T/\epsilon=1/150$ and $w/\xi= 10^4$, the last term in the brackets of Eq. (\ref{Jv}) is relatively small: 
$(T/\epsilon)\ln(w l_i/2\pi \xi^2)\simeq 0.06$. In this case the condition $J_v<J_d$ takes a simple form 
$\zeta_1\zeta_2(I_1)\lesssim 1$ implying that the control wires produce deep enough dips in $J(x)$ at the edges 
to compensate the current crowding effect of lithographic defects. Here $J_v$ decreases as $I_1$ 
increases and $\zeta_2(I_1)=J_e(I_1)/\bar{J}$ decreases so $J_v$ may drop down to $J_v\simeq J_d/2$ at $\zeta_1\zeta_2(I_1)\ll 1$, given that $\zeta_3$ 
is close to 1 in all our numerical results.

\begin{figure}[h]
\centering
   	\includegraphics[scale=0.4,trim={0mm 0mm 0mm 0mm},clip]{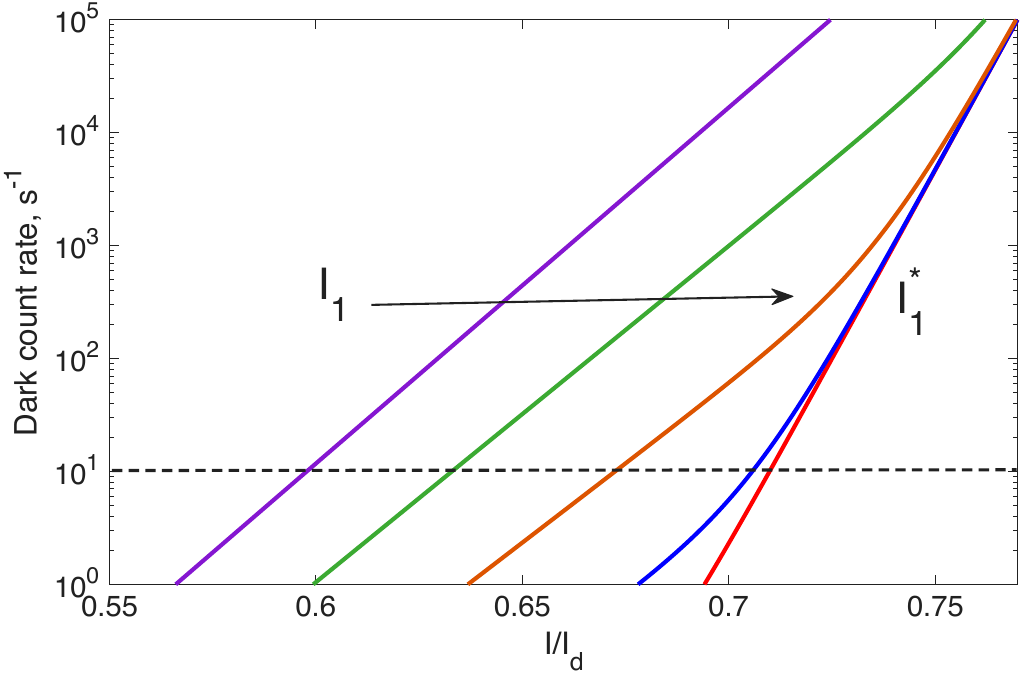}
   	\caption{Change in the logarithmic slope $d\ln S/dI$ as $I_1$ is increased calculated from Eq. (\ref{S1}) for $T=0.9$ K, $\epsilon = 66$ K~ \cite{natp}, $wl_i/\xi^2=10^4$, $S_{e0}=10^{14}$ s$^{-1}$, $\zeta_1=1.1$, $\zeta_3=1.05$ and $\zeta_2=0.9, 0.85, 0.8, 0.75, 0.7$. A jumpwise increase in $d\ln S/dI$ occurs at $\zeta_2\approx 0.7-0.75$, which defines $I_1^*$ above which $S$ is dominated by the VAV unbinding.  }
   	\label{F22}
   	\end{figure}
Shown in Fig. \ref{F22} is an example of how $d\ln S(I)/dI$ calculated from Eq. (\ref{S00}) changes upon increasing $I_1$ for the material parameters of 3 nm thick WSi detectors investigated in Ref. \onlinecite{natp}. Here the S-window is confined between the sensitivity floor (dashed line) and the upper limit close to the SNSPD switching to the normal state.  One can see that decreasing $\zeta_2$ upon increasing $I_1$ first causes a nearly parallel shift of $\ln S$ to higher bias currents followed by an abrupt increase of $d\ln I/dI$ around a threshold control current $I_1$ above which the edge vortex contribution is reduced below the sensitivity floor and $S$ is limited by the VAV pair unbinding. Furthermore, tuning $I_1$ in the control wires can reduce the dark current rate at a fixed bias current by several orders of magnitude. The behavior of $S(I,I_1)$ shown in Fig. \ref{F22} was observed on WSi SNSPD with Nb control wires ~\cite{natp}. A sharp change of $d\ln V/dI$ associated with the VAV unbinding upon increasing $I$ was observed in narrow NbN meandering nanowires ~\cite{slope}.

At $I_1>I_1^*$ penetration of vortices from the edges is suppressed and the dark count rate is limited by the VAV unbinding. In the simplest case of a flat $J(x)$ profile in the most part of the strip, Eq. (\ref{S00}) reduces to:  
\begin{gather}
S=S_{b0}\exp\left[-\frac{2\epsilon}{T}\left(1-\frac{\zeta_3 I}{I_d}\right) \right],
\label{E11}\\
S_{b0}\sim\frac{AR_\square}{8\pi^2\mu_{0}\Lambda\xi^{2}}, \qquad I_1>I_1^*.
\label{E12}
\end{gather}
The critical current $I_{c2}$ limited by the VAV pair unbinding at a given dark count rate criterion $S_c$ is then:
\begin{equation}
I_{c2}=\frac{I_{d}}{\zeta_3}\left(1-\frac{T}{2\epsilon}\ln\frac{S_{b0}}{S_{c}}\right).
\label{E13}
\end{equation}
For $A=20\,\mu$m$\times 1$ mm, $R_\square=340\,\Omega$, $\Lambda=245\,\mu$m, $\xi=7$ nm ~\cite{wsi} and $T/2\epsilon=1/300$, Eqs. (\ref{E11})-(\ref{E13}) give $S_{b0}\sim 6\cdot10^{18}\,s^{-1}$ and $I_{c2}\approx 0.87I_d/\zeta_3$ at $S_c=10^2$ s$^{-1}$. Here $I_{c2}$ decreases logarithmically with the SNSPD area.

Neither the current crowding parameter $\zeta_1$ of lithographic defects nor the depairing currents $I_d$ for a particular SNSPD are really known and they can hardly be measured directly. Usually $I_d$ is evaluated from the measurements of the kinetic inductance as a function of the bias current calculated in the BCS model  for a narrow strip with $w\ll\Lambda$ in the dirty limit ~\cite{ck}.  However, this model does not account for the realistic electron-phonon pairing of the Eliashberg theory ~\cite{eliashb}, nonequilibrium ~\cite{noneq} or Joule heating ~\cite{joule} effects, a contribution from the Josephson inductance of grain boundaries ~\cite{makita} or the reduction of $J_d$ by subgap quasiparticle states ~\cite{kubo} which are especially pronounced in thin films ~\cite{JZ}. Yet tuning $J(x)$ {\it in situ} by control wires and detecting the abrupt change in $d\ln S/dI$ at $I_1\approx I_1^*$ ensures reaching the ultimate photon sensitivity at $I=I_{c2}$ limited only by the VAV unbinding in SNSPD, even if neither $J_d$ nor the extent of current crowding at the edge defects is known.

\section{Conclusions}

This work shows that integration of a thin film superconducting strip with current-carrying control wires can be used to engineer a desired profile of supercurrent density $J(x)$ with no current crowding at the edges in a strip wider than the Pearl length. Here $J(x)$ in a strip can be tuned {\it in situ} by control wires which can produce an inverted $J(x)$ profile with controllable dips at the edges to mitigate current crowding at lithographic defects and block premature penetration of vortices.  This approach offers a principal opportunity to overcome the Pearl limit in developing {\it straight and wide} superconducting strip single photon detectors without the need of meandering SNSPDs to increase the active photon-sensitive area. Furthermore, a "flat noodle" periodic array of such straight strip detectors like those shown in Fig. \ref{F7}  can be used for the  development of multi-pixel cameras beyond the current state-of-the-art ~\cite{array}. Here each strip can be integrated with either side control wires shown in Fig. \ref{F2} or a thin film control underlayer shown in Figs. \ref{F5} and \ref{F9}, respectively. An array of bilayers tuned by its own return current shown in Fig. \ref{F7} does not require separate current sources for the underlayers and practically eliminate the Pearl current crowding in strips of any width, but do not provide controllable dips in $J(x)$ at the edges. The latter can be achieved in a strip integrated with either side controlled wires or a control underlayer fed  by separate current sources, as shown in Figs. \ref{F2} and \ref{F9}. 

The control wires can be used to tune resistive states caused by the motion of vortices. For instance, strips integrated with control wires have non-reciprocal current-voltage characteristics and can be used for the development of superconducting diodes which switch to highly resistive state above different critical currents $I_c^+$ and $I_c^-$ depending on their polarity relative to $I_1$. In turn, tuning the dips in $J(x)$ at the edges of a strip can give rise to a continuous transition from the resistance dominated by penetration of vortices from the edges to the resistance dominated by unbinding of VAV pairs. This happens as the control current exceeds a sample-dependent threshold $I_1>I_1^*$ at which an abrupt increase in the logarithmic slope of thermally-activated dark count rate $d\ln S/dI$ or voltage $d\ln V/dI$ occurs in the subcritical region of $I<I_c^+$ or $I<I_c^-$. This behavior of $S(I,I_1)$ was observed on WSi SNSPD with Nb control wires ~ \cite{natp}.   

The resistive state at $I_1>I_1^*$ is determined by the bulk VAV pair unbinding ~\cite{bkt1,bkt2}, so the measurements of the voltage-current characteristics of a strip with control wires can probe the Berezinskii-Kosterlitz-Thouless physics not masked by penetration of single vortices from the edges. Such detectors can be a unique testbed for the investigation of the extreme dynamics of superfast vortices, like those in superconducting resonator cavities ~\cite{cav}. As far as SNSPDs are concerned, they can be tuned by control wires to their ultimate  sensitivity limit because  the dark counts caused by penetration of vortices from the edges is blocked and detectors operate at the highest bias current given by Eq. (\ref{E12}), where the operational $S_c$ is only limited by the VAV pair unbinding.  This state has the highest photon sensitivity which can benefit many applications mentioned in the Introduction. 

Besides the fundamental limits of $I_{c2}$ and the photon sensitivity determined by the VAV pair unbinding, there are also technological and materials science limitations of the SNSPD performance operating at $I\simeq I_d$.  Once the width limits imposed by the Pearl screening the lithographic defects are lifted by control wires, the SNSPD performance can be limited by inhomogeneities of superconducting properties, materials defects, local non-stoichiometry, film adhesion with the substrate, etc.  Particularly grain boundaries in polycrystalline films can impede current and cause transformation of ultra-fast vortices into phase slips or VAV pairs ~ \cite{gb} at $I\ll I_d$. The current-blocking grain boundaries are characteristic of materials with short coherence length like NbN, Nb$_3$Sn, phictides or cuprates ~\cite{gbrev}, so operating SNSPDs near the depairing limit is facilitated in amorphous superconductors like WSi ~\cite{wsi}.

\acknowledgments

This work was supported by DARPA SynQuaNon Program under grant  HR0011-24-2-0386. I am grateful to  
Adam McCaughan, Marti Stevens, Kristen Parzuchowski, and Eli Mueller for many discussions and a very  
productive collaboration. I also thank Justin Allmaras and Alexei Koshelev for useful critical comments.

\appendix

\section{Numerical solution of Eqs.  (\ref{Q0}) and (\ref{Q1}).}
\label{ap1}

Following Ref. \onlinecite{natp}, Eqs. (\ref{Q0}) and (\ref{Q1}) were solved by discretizing $u_n=s(n-1)$, $n=1,... N_1$ and introducing a vector $Q_n=(Q_1,Q_2..., Q_{N_1})$ which includes $Q(sn)$ at $1<n<N$ and $Q_1(sn)$ at $N+1<n<N_1$ with $s=1/(N-1)$, $N_1=[(1+l)N]$ and $N=4\cdot 10^3$.  Here Eqs. (\ref{Q0}) and (\ref{Q1}) take the matrix form:
\begin{equation}
\sum_m M_{nm}Q_m=\alpha_n,\qquad Q_n=\sum_mM^{-1}_{nm}\alpha_m,
\label{matr}
\end{equation}
where $M_{nm}$ is given below, $\alpha_n=\alpha$ at $1<n<N$, $\alpha_n=\beta$ at $N+1<n<N_1$. To express $\alpha$ and $\beta$ in terms of currents $I=(4s/\mu_0\Lambda)\sum_{n=1}^NQ_n$ and $I_1=(2s/\mu_0\Lambda_1)\sum_{n=N+1}^{N_1}Q_n$, we sum up the second Eq. (\ref{matr}) over $n$ using that $\alpha_m$ are constants at $1<m<N$ and $N+1<m<N_1$. This yields:
\begin{gather}
I=A_0\alpha +B_0\beta,\qquad I_1=A_1\beta +B_1\alpha,
\label{equ}
\end{gather}
Hence,
\begin{gather}
\alpha=\frac{A_1I-B_0I_1}{A_0A_1-B_0B_1}, \qquad \beta=\frac{A_0I_1-B_1I}{A_0A_1-B_0B_1},
\label{sol}
\end{gather}
where
\begin{gather}
A_0=\frac{4s}{\mu_0\Lambda}\sum_{n=1}^N\sum_{m=1}^NM_{nm}^{-1},
\label{a0}\\
B_0=\frac{4s}{\mu_0\Lambda}\sum_{n=1}^N\sum_{m=N+1}^{N_1}M_{nm}^{-1},
\label{b0}\\
A_1=\frac{2s}{\mu_0\Lambda_1}\sum_{n=N+1}^{N_1}\sum_{m=N+1}^{N_1}M_{nm}^{-1},
\label{a1} \\
B_1=\frac{2s}{\mu_0\Lambda_1}\sum_{n=N+1}^{N_1}\sum_{m=1}^{N}M_{nm}^{-1}
\label{b1}
\end{gather}
The matrix elements $M_{nm}$ are given by
\begin{gather}
M_{nm}=\delta_{nm}-2ks\ln|(n^2-m^2)s^2|,
 \label{M11} \\
\quad 1<n<N,\quad 1<m<N;
\nonumber \\
\hspace{-5mm}M_{nm}=-k_1s\ln[((ms+b)^{2}-n^2s^2+h^2)^2+(2hns)^2]
\label{M12} \\
 1<n<N,\quad N+1<m<N_1; 
\nonumber \\
\hspace{-6mm}M_{nm}=-ks\ln[((ns+b)^2-m^2s^2+h^2)^2+(2hms)^2]
 \label{M21} \\
N+1<n<N_1,\quad 1<m<N;
\nonumber \\
M_{nm}=\delta_{nm} -2k_1s\ln|(ns+b)^2-(ms+b)^2|,
\label{M22} \\
 N+1<n<N_1,\quad N+1<m<N_1.
 \nonumber
\end{gather} 
The logarithmic singularity in $M_{nn}$ was integrated out: $M_{nn}\to 1- 2sk[\ln(s/2)-1+\ln(2sn)]$ and 
$M_{nn}\to 1- 2sk_1[\ln(s/2)-1+\ln(2sn+2b)]$. 

\section{Energy of the Pearl vortex in a strip}

\begin{figure}[h]
\centering
   	\includegraphics[scale=0.3,trim={20mm 40mm 0mm 40mm},clip]{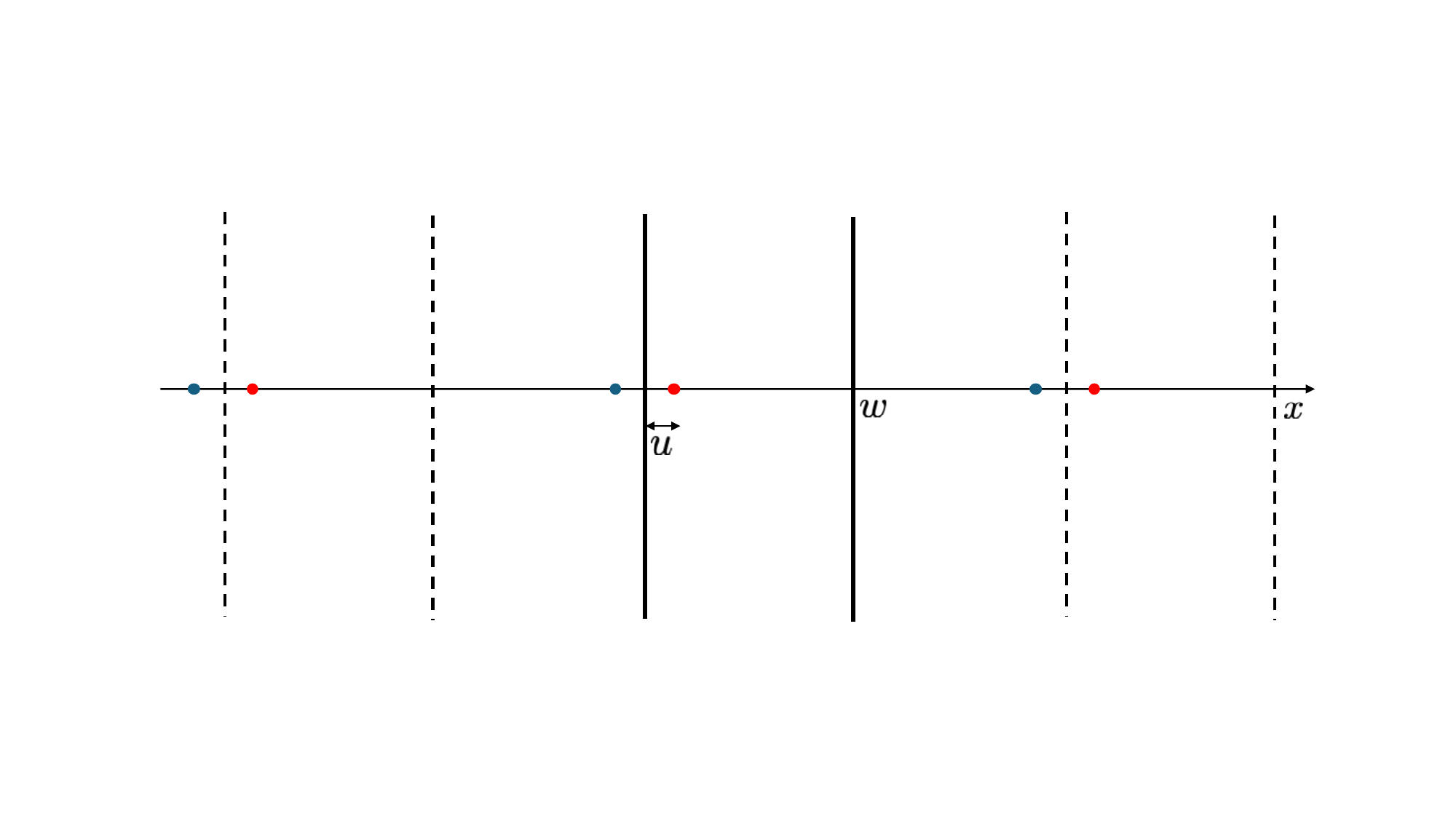}
   	\caption{Image vortices (red) and antivortices (blue) ensuring the boundary conditions $J_x(0)=J_x(w)=0$ for a vortex spaced by $u$ from the edge of a strip of width $w$.  }
   	\label{F23}
   	\end{figure}

The energy of a vortex $E(u)$ in a strip is a work against the Lorentz force of local current density at the vortex core
to move it from the edge at $x=0$ to $x=u$:
\begin{equation}
E(u)=-\phi_0\int_0^u \tilde{J}(x)dx
\label{eb1}
\end{equation}
In the London theory the driving current density at the vortex core $\tilde{J}(x)=J_i(x)+J(x)$ is a sum of the transport $J(x)$ and $J_i(x)$ induced by the edges. Following the heuristic approach of Ref. \onlinecite{gurvink}, $J_i(x,u)$ is calculated by the method of current images, as shown in Fig. \ref{F23}:
\begin{equation}
J_i=\sideset{}{'}\sum_{n=-\infty}^{\infty}J_v(x-2wn-u)-\sum_{n=-\infty}^{\infty}J_v(x-2nw+u),
\label{eb2}
\end{equation}
where $J_v(x-u)$ is the y-component of the current density produced by the Pearl vortex in an infinite film~\cite{pearl},
the first and the second sums in Eq. (\ref{eb2}) describes contributions of all image vortices and antivortices, respectively, and the prime means that the term with $n=0$ is excluded.  

Setting $x=u$ and writing Eq. (\ref{eb2}) in the Fourier space yields:
	\begin{gather}
	J_i(u)=\frac{2\phi_0}{\mu_0}\!\int \frac{d^2k}{(2\pi)^2}\frac{ik_x}{k(1+\Lambda k)}\times
	\label{eb3}\\
	\left[e^{2ik_xu}\sum_ne^{-2ik_xwn}-\sideset{}{'}\sum_ne^{-2ik_xwn}\right],
	\nonumber
	\end{gather}
where the Fourier image $J_y({\bf k})=2i\phi_0 k_x/\mu_0 k(1+\Lambda k)$ with $k^2=k_x^2+k_y^2$ for the Pearl vortex ~\cite{pearl,dg} is used. The first sum in the brackets of Eq. (\ref{eb3}) equals $(\pi/w)\sum_G\delta(k_x+G)$ with $G=\pi n/w$ and $n=0,\pm 1,\pm 2,...$. The second sum in the brackets does not contribute to $J_i(x)$ after integration over $k_x$ as all forces acting on the vortex in the strip from the image vortices cancel each other, whereas forces from the image antivortices do not  (see Fig. \ref{F23}).  Hence,
\begin{equation}
	\!\!\!J_i(u)=\frac{\phi_0}{\pi\mu_0w}\!\sum_G\int_0^\infty\!\! \frac{G\sin(2Gu)dk_y}{\sqrt{G^2+k_y^2}(1+\Lambda \sqrt{G^2+k_y^2})}
	\label{eb4}
	\end{equation}
Taking $k_y=G\tan t$ reduces the integral in Eq. (\ref{eb4}) to:
\begin{gather}
\!\!\!\int_0^{\pi/2}\!\!\!\frac{Gdt}{G\Lambda+\cos t}=
\frac{2G}{\sqrt{(G\Lambda)^2-1}}\tan^{-1}\!\!\left[\frac{G\Lambda-1} {G\Lambda+1}\right]^{1/2}
\label{eb5}
\end{gather}
As a result, $E(u)$ becomes:
\begin{gather}
E(u)=E_v(u)-\frac{2\phi_0}{\mu_0\Lambda}\int_0^uQ(x)dx,
\label{eb6}\\ 
E_v(u)=\frac{8\epsilon}{\pi}\sum_{n=1}^N\frac{\sin^2(\pi un/w)}{\sqrt{n^2-p^2}}\tan^{-1}\left(\frac{n-p}{n+p}\right),
\label{eb7}
\end{gather} 
where $p=w/\pi\Lambda$, $\epsilon=\phi_0^2/2\pi\mu_0\Lambda$ and the cutoff $N\sim w/\xi\gg 1$ is evaluated below. 
For a narrow strip $p\ll 1$, Eq. (\ref{eb7}) simplifies to:
\begin{gather}
E_v=\epsilon\sum_{n=1}^N\frac{1}{n}\left[1-\cos\frac{2\pi nu}{w}\right]=
\nonumber \\
\epsilon\left[\ln N+\gamma+\ln\left(2\sin\frac{\pi u}{w}\right)\right],
\label{eb8}
\end{gather}
where $\gamma=0.577$ is the Euler constant. The cutoff $N=[0.94w/\xi ]\gg 1$ accounts for the finite core size and the energy of the vortex core $\epsilon_0$ to be added to Eq. (\ref{eb8}):
\begin{equation}
E_v=\epsilon\ln\left[\frac{2w}{\pi\xi}\left(\sin\frac{\pi u}{w}\right)\right]+\epsilon_0
\label{eb9}
\end{equation}
Here $\epsilon_0=0.38\epsilon$ was obtained from the GL numerical calculations of $H_{c1}$ \cite{stejic} and simulations  
of a vortex in a strip~\cite{vod}. Equation (\ref{eb9}) can be written in the form~\cite{gurvink}  
\begin{equation}
E_v=\epsilon\ln\left[\frac{w}{\pi\tilde{\xi}}\sin\left(\frac{\pi u}{w}\right)\right],
\label{eb10}
\end{equation}
where $\tilde{\xi}=0.34\xi$ and $0.34=0.5\exp(-0.38)$ absorbs the numerical factors, so that $E_v(u)\to 0$ at $u=\tilde{\xi}\ll w$. 
To improve convergence of the series in Eq. (\ref{eb7}), Eq. (\ref{eb10}) was added and subtracted in Eq. (\ref{euu}).

\end{document}